\documentclass[fleqn,usenatbib,linenumbers]{mnras}
\usepackage{newtxtext,newtxmath}

\usepackage{lineno}
\usepackage[T1]{fontenc}
\usepackage{chngcntr}
\usepackage{appendix}
\usepackage{tablefootnote}
\usepackage{caption}
\usepackage{footnote}
\makesavenoteenv{tabular}
\DeclareRobustCommand{\VAN}[3]{#2}
\let\VANthebibliography\thebibliography
\def\thebibliography{\DeclareRobustCommand{\VAN}[3]{##3}\VANthebibliography}

\usepackage{threeparttable}
\usepackage{graphicx}	
\usepackage{amsmath}	
\usepackage{amssymb}	
\usepackage{booktabs}

\newcommand{\HII}{\mathrm{H}\,\textsc{\large{ii}}}

\newcommand{\tco}{$^{13}$CO(3-2)}
\newcommand{\cetno}{C$^{18}$O(3-2)}
\newcommand{\kms}{\,km\,s$^{-1}$}

\usepackage{scalerel,tikz}
\usetikzlibrary{svg.path}
\definecolor{orcidlogocol}{HTML}{A6CE39}
\tikzset{orcidlogo/.pic={
 \fill[orcidlogocol] svg{M256,128c0,70.7-57.3,128-128,128C57.3,256,0,198.7,0,128C0,57.3,57.3,0,128,0C198.7,0,256,57.3,256,128z};
 \fill[white] svg{M86.3,186.2H70.9V79.1h15.4v48.4V186.2z}
 svg{M108.9,79.1h41.6c39.6,0,57,28.3,57,53.6c0,27.5-21.5,53.6-56.8,53.6h-41.8V79.1z M124.3,172.4h24.5c34.9,0,42.9-26.5,42.9-39.7c0-21.5-13.7-39.7-43.7-39.7h-23.7V172.4z}
 svg{M88.7,56.8c0,5.5-4.5,10.1-10.1,10.1c-5.6,0-10.1-4.6-10.1-10.1c0-5.6,4.5-10.1,10.1-10.1C84.2,46.7,88.7,51.3,88.7,56.8z};
}}
\newcommand\orcidicon[1]{\href{https://orcid.org/#1}{\mbox{\scalerel*{
\begin{tikzpicture}[yscale=-1,transform shape]
\pic{orcidlogo};
\end{tikzpicture}
}{|}}}}

\graphicspath{{plots/}}

\title[Study of N59]{Unveiling the Cosmic Cradle: clustering and massive star formation in the enigmatic Galactic bubble N59}

\author[S. T. Paulson, K. K. Mallick \& D.K. Ojha]{
Sonu Tabitha Paulson$^{\orcidicon{0000-0002-8131-7020}}$,$^{1}$\thanks{E-mail: sonu.paulson@tifr.res.in}
K. K. Mallick$^{\orcidicon{0000-0002-3873-6449}\,2}$ and 
D. K. Ojha$^{\orcidicon{0000-0001-9312-3816}\,1}$
\\
$^{1}$Tata Institute of Fundamental Research, Mumbai (Bombay) 400005, India\\
$^{2}$Aryabhatta Research Institute of Observational Sciences (ARIES), Nainital 263129, India\\
}

    \date{Accepted 2024 March 27. Received 2024 March 12; in original form 2023 September 27}

\pubyear{2024}

\begin{document}
\label{firstpage}
\pagerange{\pageref{firstpage}--\pageref{lastpage}}
\maketitle
\begin{abstract}
In this paper, we have conducted an investigation focused on a segment of the \emph{Spitzer} mid-infrared bubble N59, specifically referred to as R1 within our study. Situated in the inner Galactic plane, this region stands out for its hosting of five 6.7 GHz methanol masers, as well as numerous compact $\HII$~regions, massive clumps, filaments, and prominent bright rims. As 6.7 GHz masers are closely linked to the initial phases of high-mass star formation, exploring regions that exhibit a high abundance of these maser detections provides an opportunity to investigate relatively young massive star-forming sites. To characterize the R1 region comprehensively, we utilize multi-wavelength (archival) data from optical to radio wavelengths, together with $^{13}$CO and C$^{18}$O data. Utilizing the \emph{Gaia} DR3 data, we estimate the distance towards the bubble to be $4.66 \pm 0.70$ kpc. By combining near-infrared (NIR) and mid-infrared (MIR) data, we identify 12 Class I and 8 Class II sources within R1. Furthermore, spectral energy distribution (SED) analysis of selected sources reveals the presence of four embedded high-mass sources with masses ranging from 8.70-14.20~M$_\odot$. We also identified several O and B-type stars from radio continuum analysis. Our molecular study uncovers two distinct molecular clouds in the region, which, although spatially close, occupy different regions in velocity space. We also find indications of a potential hub-filament system fostering star formation within the confines of R1. Finally, we propose that the feedback from the $\HII$ regions has led to the formation of prominent Bright Rimmed Clouds (BRC) within our region of interest.

\end{abstract}


\section{INTRODUCTION}\label{R1}

Massive stars (> 8 M$_\odot$), characterized by their substantial mass and intense luminosity, hold a pivotal role in various astrophysical phenomena. Their significance spans several domains, from contributing to the synthesis of heavy elements and enriching the interstellar medium (ISM) \citep{zinnecker2007toward, motte2018high}, to influencing the dynamics of their host galaxies through their radiation and stellar winds, thereby shaping gas motions and fostering the birth of new stars. Despite their important role, the study of massive star formation presents challenges due to their rarity, rapid evolution, and their concealment within dusty cocoons during their early stages. 

One promising avenue for probing massive star-forming regions involves the detection of 6.7 GHz methanol masers, as these are closely associated with the initial phases of massive star formation \citep{breen2013confirmation, paulson2020probing}. In this study, we focus our attention on the specific star-forming region known as MIR bubble N59. This region stands out for hosting approximately eight 6.7 GHz methanol masers, alongside numerous compact $\HII$ regions, massive clumps, filaments, and prominent bright rims. Analyzing the characteristics of this region, therefore, allows us to glean valuable insights into the interplay between the processes involved in the early phase of high-mass star formation.


The Galactic mid-infrared (MIR) bubble N59, initially documented by \cite{churchwell2006bubbling}, is situated at coordinates (l,b)= (33.071$^\circ$,-0.075$^\circ$). 
\citet{deharveng2010gallery} offer a concise description of bubble N59, highlighting the presence of multiple bright condensations and radio sources identified from the catalogues of \citet{becker19945} and \citet{kurtz2005water}. They also report the identification of two 6.7 GHz methanol masers within the confines of the bubble, referencing compilations from \citet{pestalozzi2005general} and \citet{xu2008high}. Notably, these masers are situated in the vicinity of bright dust condensations that are interacting with $\HII$ regions, implying that N59 holds significant potential as a candidate for investigating the mechanisms driving triggered massive star formation. \citet{hattori2016mid} have also targeted bubble N59 as part
of their sample of 180 IR bubbles for a statistical study of mid- and far-infrared properties of \textit{Spitzer} Galactic bubbles. They find N59 to be a closed bubble that hosts polycyclic aromatic hydrocarbons (PAH) with luminosities exceeding 10$^5$~L$_\odot$. 
Subsequently, \citet{hanaoka2019systematic} conducted a follow-up study involving 247 IR bubbles, including N59, incorporating data from the \emph{Herschel} survey. The investigation revealed that the fractional PAH luminosity (i.e. L$_{\rm PAH}$/L$_{\rm TIR}$) of N59 was approximately 0.17. 
Based on the L$_{\rm TIR}$ value \citep{Martins2005}, presence of O-type stars was proposed as one of the reasons for the low L$_{\rm PAH}$/L$_{\rm TIR}$, as the intense UV luminosities from such stars would enhance L$_{\rm TIR}$ and expedite the photodissociation of PAH molecules. Thus, N59 turns out to be harbouring several bright stars, including massive stars. While bubble N59 has been featured in several statistical investigations concentrating on dust properties, it has not yet been subjected to a comprehensive multiwavelength analysis. Given the intriguing findings in previous research, such as distinctive dust distribution, the presence of 6.7~GHz methanol masers near prominent dust condensations, and indications of potential triggered star formation, this region merits an in-depth examination.

\begin{figure*}
\includegraphics[width=0.45\textwidth, trim= 0 0.1cm 0 0cm]{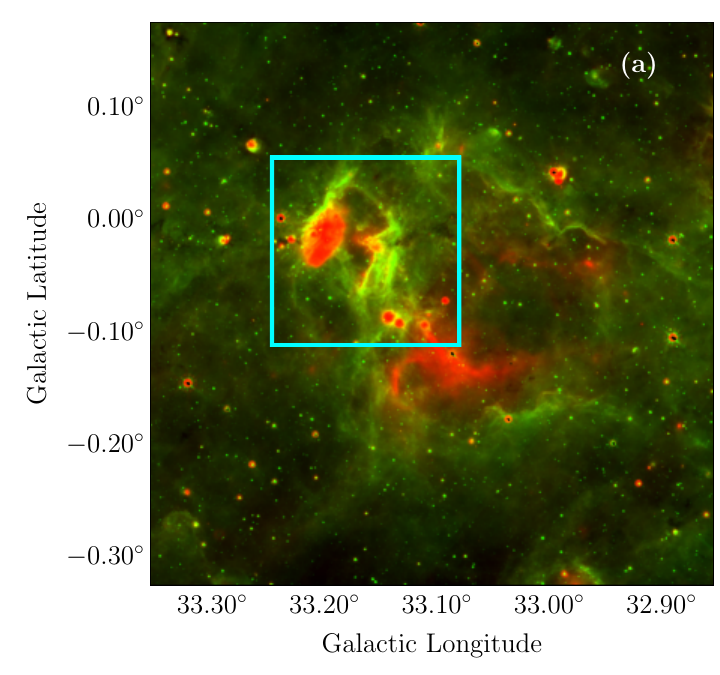}\quad
\includegraphics[width=0.44\textwidth, trim= 0 0.1cm 0 0cm]{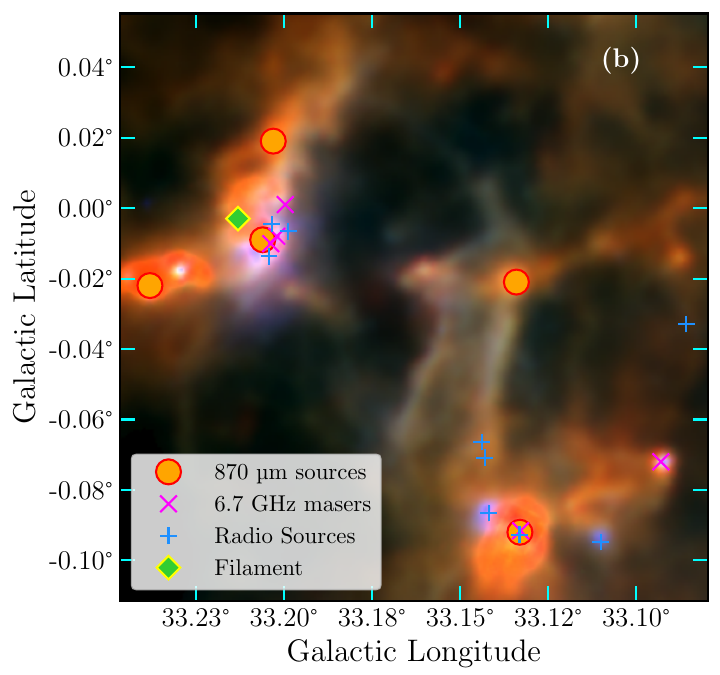}

\caption{Large-scale view of the MIR bubble N59 (size of the selected ﬁeld is 30\arcmin×30\arcmin with central coordinates of \ensuremath{l=33.071}$^\circ$; \ensuremath{b=-0.075}$^\circ$). Left panel (a) A two-color-composite map (\emph{Spitzer}-MIPSGAL 24~\textmu m (red) and \emph{Spitzer}-GLIMPSE 8.0~\textmu m (green) images. The cyan square marked is region R1. Right panel (b) Three colour composite map of region R1 using \emph{Herschel} data (250~\textmu m  (red), 160~\textmu m  (green), and 70~\textmu m  (blue) images in linear scale). Overlaid are 870~\textmu m ATLASGAL clumps (orange circles), 6.7 GHz maser emissions (magenta crosses), radio sources (blue plusses) and the intersection point with the filament identified by \citet{li2016atlasgal} (marked as a green diamond).}
\label{multiview}
\end{figure*}

Fig.~\ref{multiview}(a) displays a two-color composite image of N59, showcasing the dust emission at 8.0~\textmu m (green) and 24.0~\textmu m (red) from \emph{Spitzer}-MIPSGAL. The bubble stands out prominently within this composite map, spanning an area of $\sim $30\arcmin×30~\arcmin. It is worth highlighting that a significant concentration of radio sources, ATLASGAL clumps \citep{schuller2009atlasgal}, and methanol masers is primarily situated within the cyan square measuring 10' X 10' (hereafter referred to as 'R1').

An enlarged view of R1 is provided in Fig.~\ref{multiview}(b), which is a three-color-composite map created using the \emph{Herschel} data, where the red, green and blue components correspond to 250~\textmu m, 160~\textmu m and 70~\textmu m, respectively, for region R1. The map also features the position of Galactic filament (marked using the green diamond) reported by \cite{li2016atlasgal}, who discerned it by employing the DisPerSE algorithm \citep{sousbie2011persistent} on the ATLASGAL survey data. The magenta crosses represent 6.7~GHz Class II methanol maser detections. Additionally, the map reveals the presence of multiple bright condensations that coincide with 870~\textmu m ATLASGAL clumps. Numerous radio sources (the blue plus symbols; see Section~\ref{radiodata}) are concentrated within the bubble, primarily around the brightest condensations. The 6.7~GHz methanol masers in R1 were identified using the online tool, \emph{MaserDB}\footnote{https://maserdb.net/}, provided by \citet{ladeyschikov2019online}. 
As the majority of ATLASGAL sources, $\HII$ regions, and methanol maser detections are concentrated in R1, our study focuses extensively on this specific area.

\begin{table*}
\centering
\caption{Catalog of various surveys endorsed for this study, covering a wide range of wavelengths from near-infrared to radio frequencies.}
\begin{threeparttable}
\scriptsize
\begin{tabular}{c c c c}
\hline
Survey & Wavelength/s & $\sim$ Resolution & Reference\\
\hline
\emph{Herschel} Infrared Galactic Plane Survey$^{\rm a}$& 70, 160, 250, 350, 500 $\mu$m & $5\arcsec.8$, $12\arcsec$, $18\arcsec$, $25\arcsec$, $37\arcsec$ & \citet{molinari2010hi}\\
NRAO VLA Sky Survey (NVSS)$^{\rm b}$ & 21 cm & $46\arcsec$ & \citet{condon1998nrao}\\
Multi-Array Galactic Plane Imaging Survey (MAGPIS)$^{\rm c}$ & 20 cm & $6\arcsec$ & \citet{helfand2006magpis}\\
Co-Ordinated Radio `N' Infrared Survey for High-mass star formation (CORNISH)$^{\rm d}$ & 6 cm & $1.5\arcsec$ & \citet{hoare2012coordinated}\\
VLA Archive$^{\rm e}$  &8.49 GHz    &9\arcsec &\citet{crossley2007nrao} \\
Galactic Legacy Infrared Mid- plane Survey Extraordinaire (GLIMPSE): \emph{Spitzer}-IRAC$^{\rm f}$ &3.6, 4.5, 5.8, and 8.0 $\mu$m  & $\sim$2$\arcsec$, $\sim$2$\arcsec$, $\sim$2$\arcsec$, $\sim$2$\arcsec$ &\citet{benjamin2003glimpse} \\
The APEX Telescope Large Area Survey of the Galaxy (ATLASGAL)$^{\rm g}$ & 870~$\mu$m &$36\arcsec$ & \citet{schuller2009atlasgal}\\
The UKIRT Infrared Deep Sky Survey (UKIDSS)$^{\rm h}$ & J (1.2~$\mu$m), H (1.6~$\mu$m) and K (2.2~$\mu$m) &$0.8\arcsec$, $0.8\arcsec$, $0.8\arcsec$ & \citet{lawrence2007ukirt}\\
Global Astrometric Interferometer for Astrophysics (GAIA)$^{\rm i}$ &330-1050 nm &$0.4$ mas&\citet{vallenari2023gaia} \\ 
CO Heterodyne Inner Milky Way Plane Survey (CHIMPS)$^{\rm j}$  &0.09 cm &$15\arcsec$ &\citet{Rigby_CHIMPS_2016MNRAS} \\
\hline
\end{tabular}
\begin{tablenotes}
\item[] $^\mathrm{a}$http://archives.esac.esa.int/hsa/whsa/; $^\mathrm{b}$https://www.cv.nrao.edu/nvss/postage.shtml; $^\mathrm{c}$https://third.ucllnl.org/gps/; $^\mathrm{d}$https://cornish.leeds.ac.uk/public/index.php; $^\mathrm{e}$https://science.nrao.edu/facilities/vla/archive/; $^\mathrm{f}$https://irsa.ipac.caltech.edu/cgi-bin/Gator/nph-dd; $^\mathrm{g}$https://www3.mpifr-bonn.mpg.de/div/atlasgal/; $^\mathrm{h}$http://www.ukidss.org/; $^\mathrm{i}$https://gea.esac.esa.int;\qquad $^\mathrm{j}$https://www.canfar.net/en/.
\end{tablenotes}
\end{threeparttable}
\label{table_DataSet}
\end{table*}


\section{Archival Data}\label{S2}
Table~\ref{table_DataSet} presents a summary of the archival data used in this paper. The subsequent portion of this Section offers a concise description of each data sets.

\subsection{\emph{Gaia} DR3}
The third \emph{Gaia} data release (\emph{Gaia} DR3) contains the positions, parallaxes, proper motions, and apparent brightness in the G-band ($0.33-1.05$ \textmu m) for about 1.8 billion stars, based on 34 months of mission observations \citep[\emph{Gaia} Collaboration,][]{vallenari2023gaia}. Incorporated within \emph{Gaia} DR3 are measurements of the apparent brightness in the G$_{BP}\,(0.33-0.68$ \textmu m) and G$\mathrm{_{RP}\,(0.63-1.05}$ \textmu m) bands, providing a wealth of broad-band color information. We used the \emph{Gaia} data to retrieve the proper motion of stars as well as to determine the distance of the complex. The data was downloaded from the \emph{Gaia} archive \footnote{https://gea.esac.esa.int/archive}.


\subsection{Near-Infrared Data from UKIDSS}
The near-infrared (NIR) J (1.25 \textmu m), H (1.65 \textmu m) and K (2.16 \textmu m) data are retrieved from the UKIRT Infrared Deep Sky Survey (UKIDSS) DR10PLUS Galactic Plane Survey \citep[GPS;][]{lawrence2007ukirt}. The UKIDSS data is obtained using the Wide Field Camera (WFCAM) on the United Kingdom Infrared Telescope. The UKIDSS data is publicly available in WFCAM science archive. Following the selection procedure described in \cite{lucas2008ukidss} and \cite{dewangan2015physical}, we downloaded the NIR data using the Structured Query Language (see Appendix~\ref{sql}). In order to ensure reliable photometry, we discarded the sources that have photometric uncertainty > 0.1 magnitude in each band. 

\subsection{Mid-Infrared Data from \emph{Spitzer}-IRAC}
We obtained the archival mid-infrared (MIR) data using the \emph{Spitzer} Space Telescope observations under the Galactic Legacy Infrared Mid-Plane Survey Extraordinaire (GLIMPSE) program \citep[][]{benjamin2003glimpse,churchwell2009spitzer}. GLIMPSE is a legacy science program that mapped the inner Galactic plane using the Infrared Array Camera (IRAC) across four distinct bands: 3.6 (I1), 4.5 (I2), 5.8 (I3), and 8.0 (I4) microns. The angular resolution in the four bands is less than 2\arcsec \, \citep{fazio2004number}. The GLIMPSE point sources were retrieved from the highly reliable GLIMPSE 1 Spring '07 archive \footnote{https://irsa.ipac.caltech.edu/applications/Gator/}.

\subsection{Far Infrared and Sub-millimeter Data}
The far-Infrared (FIR) continuum images are retrieved from the \emph{Hershel} Space Observatory data archives.  We made use of the processed level2\_5 images at 70, 160~\textmu m (PACS) as well as 250, 350 and 500~\textmu m (SPIRE), observed as a part of the \emph{Herschel} infrared Galactic plane survey \citep[Hi-GAL;][]{molinari2010hi}. The maps at different wave bands have different data units, resolution and plate scales. The
70 \textmu m, 160 \textmu m, 250 \textmu m, 350 \textmu m, and 500 \textmu m images have pixel
scales of 3.2\arcsec, 6.4\arcsec, 6\arcsec, 10\arcsec, and 14\arcsec, respectively, with resolutions varying from $\sim$ 5.5\arcsec$-$ 36\arcsec. While the PACS images obtained had the surface brightness unit of Jy pixel$^{-1}$, the SPIRE images were in the units of MJy sr$^{-1}$. The sub-millimeter continuum map at 870~\textmu m (beam size $\sim$ 19.2\arcsec) was retrieved from the ATLASGAL archival survey \citep{schuller2009atlasgal}. The FIR and sub-millimeter data are employed to analyze the prevailing physical conditions within the region. 

\subsection{Column density and Temperature Maps}\label{vialectea}
The \emph{Herschel} column density and temperature maps (spatial resolution$\sim$ 12\arcsec) have been downloaded directly from the publicly available website\footnote{http://www.astro.cardiff.ac.uk/research/ViaLactea/}. These maps are procured for EU-funded ViaLactea project \citep{molinari2010hi} adopting the Bayesian PPMAP technique \citep{molinari2010clouds} at 70, 160, 250, 350, and 500 \textmu m wavelengths \emph{Herschel} data \citep{marsh2015temperature,marsh2017multitemperature}. 

\subsection{Radio Continuum Data} \label{radiodata}
For obtaining the radio continuum image of the region, we have used the radio interferometer surveys of the Galactic Plane, namely CORNISH \citep[5 GHz;][]{hoare2012coordinated}, MAGPIS \citep[1.4 GHz;][]{helfand2006magpis}, and NVSS \citep[1.4 GHz;][]{condon1998nrao}. The angular resolution for CORNISH, MAGPIS and NVSS surveys are 1.5\arcsec, 6.0\arcsec and 45\arcsec respectively. 
While CORNISH and MAGPIS concentrate on compact sources (at high resolution and low resolution respectively), NVSS excels in capturing extended emissions \citep{10.1093/mnras/stz347}. We also obtained archival VLA data at 8.49~GHz for a small part of our R1 region 
from the NRAO VLA Archive Survey (NVAS) Images Page\footnote{https://www.vla.nrao.edu/astro/nvas/} (Obs Date : 1997 Nov 22, Version : 2007 July 10).


\subsection{Molecular CO Data}
The J= 3$-$2 line of the $^{13}$CO data is obtained from the CO Heterodyne Inner Milky Way Plane Survey \citep[CHIMPS;][]{Rigby_CHIMPS_2016MNRAS}. This high-resolution spectral survey  has been carried out using the Heterodyne Array Receiver Program on the 15 m James Clerk Maxwell Telescope (JCMT) in Hawaii. CHIMPS has an angular resolution of 15\arcsec~in 0.5 km s$^{-1}$ velocity channels. 

\section{RESULTS}\label{Res}

\begin{figure}
\begin{center}
\includegraphics[width=0.4\textwidth, trim= 0cm 0.1cm 0 0cm]{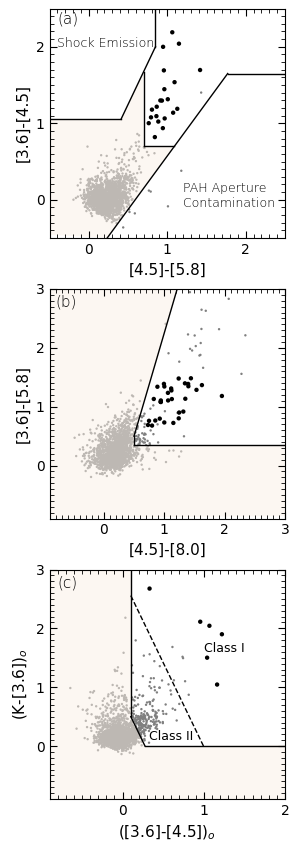}
\caption{The Class I and Class II YSOs (filled black circles) identified using CC diagrams. Panels (a) and (b) represent Class I and Class II sources, respectively, obtained using the Phase 1 scheme \citep{gutermuth2009spitzer}. Panel (c) shows Class I and Class II sources in the NIR/IRAC CC plot with areas demarcated (by black dashed and solid lines) as per the Phase 2 scheme. The filled black circles mark the additionally identified Class 0/I sources. Grey dots within both shaded and unshaded regions across panels represent unclassified sources, and sources categorized as contaminants.} \label{fig:MIRclass}
\end{center}
\end{figure}
\begin{figure}
\begin{center}
W\includegraphics[width=0.45\textwidth, trim= 0 0.1cm 0 0]{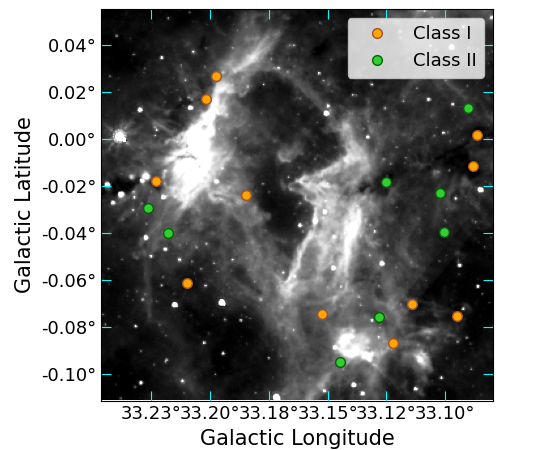}
\caption{Class I (orange filled circles) and Class II sources (green filled circles) identified using \emph{Spitzer} and UKIDSS data \citep{gutermuth2009spitzer}. The YSOs are overlaid on \emph{Spitzer}-GLIMPSE 8~\textmu m image.} \label{fig:c1c2160}
\end{center}
\end{figure}

\subsection{Notes on Distances} \label{dist}

An accurate determination of the distance to the star-forming region R1 is essential for estimating parameters like the size and mass of the associated dust condensations. \cite{deharveng2010gallery} report a kinematic distance of 5.6 kpc for N59, derived primarily from the velocity measurements of the ionized gas using radio recombination lines. This calculation assumes circular rotation around the Galactic center. However, the bubble 
lies within the ``Solar Circle'' (i.e. $\lesssim8$~kpc) and therefore carries a distance ambiguity due to two potential solutions (e.g. \citealt{kolpak2003resolving} and references therein). This can be avoided by using the distances derived from parallax.

Thanks to the new \emph{Gaia} data release (DR3; \citealt{vallenari2023gaia}), we now have precise parallax measurements for over 1.8 billion stars, which can be leveraged to derive robust distance estimates. The inversion of parallax to distance is, however, non-trivial, mainly due to the positivity constraint of distance and the non-linear transformation \citep{Luri2018}. In fact, reliable distances cannot be obtained for the vast majority of stars through this inversion \citep[see the discussion by][]{bailer2018,bailer2021estimating}. As such, the most appropriate way to implement this conversion is through a probabilistic approach. In this regard, \citet{bailer2021estimating} present a promising method that uses parallax to provide the full likelihood distribution of the distance. These distances have even been shown to carry smaller uncertainties than kinematic distances, especially for the stars situated at distances less than 5 kpc \citep{karska2022census}. Considering these advantages, we chose to utilize the stellar distances provided by \citet{bailer2021estimating}.

{To this end, we considered all stars belonging to region R1 (i.e, we are not applying the clustering algorithm on any pre identified cluster), with parallax uncertainties below 10 per cent, and computed their membership probabilities using the Unsupervised Photometric Membership Assignment in Stellar Clusters algorithm (UPMASK).\footnote{This is implemented using \textsc{py}\textsc{upmask} \citep{pera2021pyupmask}, which requires RA, Dec, parallax and proper motion (along RA and Dec) as inputs, and returns the clustering probabilities.} The algorithm has the advantage of being non-parametric and unsupervised, that is, no a priori selection of field stars is required to serve as a comparison model -- which is generally the case in many of the previous cluster membership calculators. Based on the values thus obtained, we selected the stars with membership probabilities $\geq$ 80 per cent, and used them to infer the mean distance to R1: $4.66 \pm 0.70$ kpc. More details on calculating membership probabilities using PyUPMASK are provided in Appendix~\ref{pyup}


\begin{table*}
\small
\centering
\caption{Main parameters from SED analysis.}
\begin{tabular}{c c c c c c c c c}
\hline
\emph{l} &\emph{b} & Class & $\mathrm{\chi^{2}}$ &A$\mathrm{_v}$ & Mass &log (Age) &log (Luminosity) &log (Mass accretion rate)\\
(deg) &(deg) & &  & & (M$_\odot$) &(yr) &(L$_\odot$) &(M$_\odot$ yr$^{-1}$)\\
\hline
33.086 &+0.002 &Class I   &1.15  &5.25   &14.2  &3.33  &3.82 & -3.23 \\
33.088 &-0.012 &Class I   &0.03  &7.62  &8.74  &3.77  &3.30  &-3.60 \\
33.198 &+0.027 &Class I   &0.13  &9.04  &5.32  &3.56  &2.62 &-4.35 \\
33.185 &-0.024 &Class I   &0.01  &3.98  &2.77  &4.9   &1.80  &-4.77 \\
33.095 &-0.076 &Class I   &0.32  &4.88  &8.7   &4.09  &2.96  &-3.75\\
33.223 &-0.018 &Class I   &0.37  &8.93  &4.91  &5.69  &2.21  &-5.87\\
33.210 &-0.061 &Class I   &0.14  &4.39  &8.74  &3.77  &3.30  &-3.60 \\
33.090 &+0.013  &Class II  &0.02  &9.17  &4.25  &5.57  &1.69 &-5.94 \\
33.102 &-0.023 &Class II  &0.03  &9.01  &4.12  &5.87  &1.87 &-5.84 \\
33.125 &-0.018 &Class II  &2.50   &0.10  &3.97  &4.99  &2.01 &-4.44 \\
33.100 &-0.040   &Class II  &0.01  &3.88  &3.82  &5.58  &1.52 &-5.81  \\
33.128 &-0.076 &Class II  &0.03  &7.47   &7.88  &3.85  &3.08 &-3.72 \\
33.227 &-0.029 & Class II &0.79  &0.78   &7.63  &5.16  &3.30  &-3.48  \\
33.145 &-0.095 & Class II &2.66  &1.07   &2.98  &4.82  &1.86  &-3.46 \\              
\hline
\end{tabular}
\label{table_SEDparams}
\end{table*}

\subsection{Young Stellar Population}\label{S3}
The evolutionary state of a young stellar object (YSO) is often signposted by its infrared color excess - an indicator of ionized stellar wind or circumstellar dust \citep{deng2022infrared}. The infrared color-color (CC) and color-magnitude (CM) diagrams display distinct and specific regions occupied by various object types, indicating that each type is well-defined. Additionally, the position of a young stellar object in CC/CM space can provide some valuable information about its evolutionary stage. We utilize the GLIMPSE 1 and UKIDSS data to investigate the infrared excess sources present in R1. However, the selection of YSOs based on excess emission in the infrared is easily contaminated by external polycyclic aromatic hydrocarbon (PAH)/star-forming galaxies, active galactic nuclei (AGNs), or shock-excited extended sources, as they often mimic YSO colors \citep{gutermuth2009spitzer}. Hence, we eliminated these contaminants prior to classifying the YSOs, using the methods described in \cite{gutermuth2009spitzer}. The following steps were adopted to identify and classify YSOs:

\begin{enumerate}
\item IRAC I1-I2-I3-I4 catalog: Initially, the YSOs were identified using their MIR magnitudes. For this purpose, we obtained the \emph{Spitzer} sources from the GLIMPSE 1 catalog within a search radius of 9~\arcmin centered at (l,b) = (33.068$^\circ$, -0.044$^\circ$). We ensured that the errors in each individual band are $\leq$ 0.15 mag. This generated a comprehensive catalog of 2242 sources. We employed the Phase 1 method described in \citet[Appendix A.1]{gutermuth2009spitzer} for YSO identification. The YSOs are classified into Class I (protostars with circumstellar disks and infalling envelops) and Class II (pre-main sequence stars with optically thick disks), based on their position in the  [4.5]-[5.8] Vs [3.6]-[4.5] and [4.5]-[8.0] Vs [3.6]-[5.8] CC diagram, respectively. We obtained a total of 22 Class I and 31 Class II sources using this method, which are shown in Fig.~\ref{fig:MIRclass} (a) and (b). Among this, 11 Class I and 8 Class II sources fall in R1.
\item H-K-I1-I2 catalog: There have been cases where there were non-detections at 5.8 and 8.0 \textmu m, but good quality  detections in the NIR bands. We used a combination of H, K (from UKIDSS survey data), and \emph{Spitzer}-GLIMPSE I1, I2 data, for classifying the YSOs in such scenarios. The H, K, and I1, I2 sources were cross-matched with a 0.6\arcsec matching radius, resulting in a catalog containing 4079 sources. Using the classification scheme described in  \citet[see their Appendix A Phase 2]{gutermuth2009spitzer}, the YSOs were identified from their location in the dereddened (K-[3.6]) Vs [3.6]-[4.5] CC diagram. The dereddened colors were obtained using the color excess ratio $\mathrm{\frac{E_{H-K}}{E_K-[3.6]} = 1.49}$ \citep{flaherty2007infrared}. We obtained an additional 6 Class 0/I sources using this method (see Fig.~\ref{fig:MIRclass}(c)) with one of them being associated with region R1.

In summary, within region R1, we detected 11 Class I and 8 Class II sources exclusively using the IRAC catalog, and an additional Class I source by combining data from both NIR and IRAC catalogs. This brings the total count of identified YSOs in R1 to 12 Class I and 8 Class II sources. The locations of these identified YSOs are shown in Fig.~\ref{fig:c1c2160}.
\begin{figure}
\begin{center}
\includegraphics[width=0.5\textwidth, trim= 0 0.1cm 0 0]{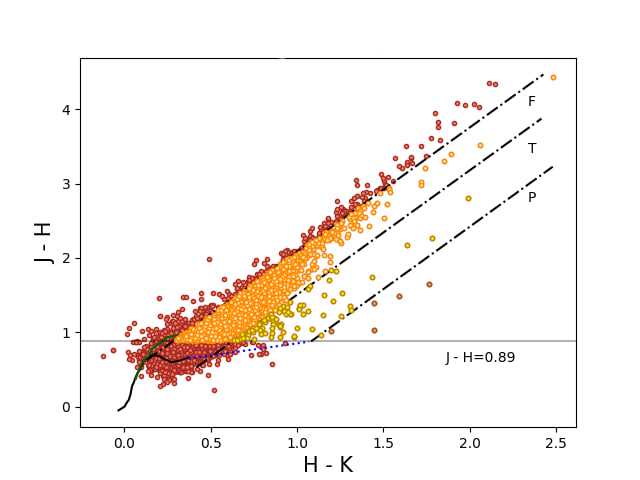}
\caption{J - H/H - K CC diagram for the region. The black curve marks the locus of the dwarfs from \citet{hewitt2006green} The dotted blue line is the Classical T Tauri star (CTTS) locus from \citet{yasui2008star}. The dot–dashed black lines are the reddening vectors, drawn using the reddening laws from \citet{Rieke1985}. Three regions-"F," "T," and "P"-mark the location of different classes of YSOs, with sources in each region colored in orange, yellow and brown, respectively. The horizontal gray line has been drawn at J - H = 0.89. All points are in the Mauna Kea Observatory (MKO) system.} \label{fig:NIRclass}
\end{center}
\end{figure}
\begin{figure}
\begin{center}
\includegraphics[width=0.5\textwidth, trim= 0 0.1cm 0 0]{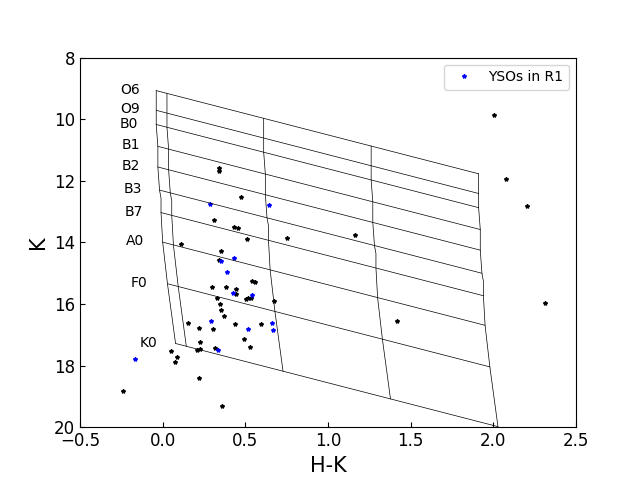}
\caption{K/H - K NIR CM-D for bubble N59. Nearly vertical solid lines are the loci of ZAMS stars reddened by A$_{\rm V}$ = 0, 1, 10, 20 and 30 mag. Parallel, slanting lines indicate the reddening vectors for respective spectral types. The small star symbols represent the total YSOs in bubble N59, with those belonging to region R1 colored blue. All points are in the Mauna Kea Observatory (MKO) system.} \label{fig:KHK}
\end{center}
\end{figure}

\end{enumerate}


\subsubsection{Spectral Energy Distribution of YSOs}\label{yso}
We conducted spectral energy distribution (SED) modeling for a subset of 12 Class I and 8 Class II YSOs, to estimate their physical parameters. To accomplish this, we employed the grid of YSO models developed by \cite{robitaille2006interpreting} and utilized the online SED fitting tool created by \cite{robitaille2007interpreting}. The underlying model includes a pre-main sequence (PMS) star surrounded by a flared accretion disk, which is enveloped by a rotationally flattened envelope featuring cavities carved out by a bipolar outflow. The SED models were computed using the radiation transfer code of \cite{whitney2003two,whitney2004glimpse} in a 14-dimensional parameter space, encompassing properties of the central source, infalling envelope, and disk. In the fitting process, a total of 200,000 SED models were generated, and each fitting was characterized by a $\chi^2$ parameter, comparing the available SED models with the data. The range for distance and interstellar visual extinction (A$\mathrm{_V}$) were considered as free parameters, with a distance range of 4.5 to 5.2 kpc for our sources.

To quantify the extinction, we employ a method that involves measuring the color excess in the infrared wavelengths, as described by \citet{Lada1994}. In our study, we utilized the NIR J - H/H - K CC diagram of the region R1 to identify sources displaying reddening. We specifically focused on sources within the "F" region (refer to Fig.~\ref{fig:NIRclass}) that were not initially classified as YSOs. We only considered those sources situated above the J - H = 0.89 threshold as a measure to reduce potential contamination. To estimate the average visual extinction in this region, we adapted a method similar to the Near-IR-Color-Excess technique proposed by \citet{Lada1994}, \citet{Ojha2004} and \citet{Kainulainen2007}. Initially, we calculated the color excess (H - K) for each selected sources by dereddening them along the reddening vectors until they aligned with the approximate straight-line locus representing dwarf stars. Subsequently, A$_{\rm V}$ was computed for each source using the reddening laws established by \citet{Rieke1985}. The A$_{\rm V}$ values lie in the range of 2.5-9~mag. Furthermore, we plot the K/H-K colour–magnitude diagram (CM-D) for the entire bubble (refer Fig.~\ref{fig:KHK}). The almost vertical solid lines represent the zero-age main sequence (ZAMS) loci at a distance of 4.66~kpc reddened by A$_{\rm V}$ = 0, 1, 10, 20 and 30 mag. Slanting lines indicate the reddening vectors for the marked spectral types. The 59~YSOs obtained (see sect.~\ref{S3}) are depicted as tiny star symbols, with those belonging to region R1 highlighted in blue. Notably, a majority of our YSOs are distributed between A$_{\rm V}$ = 1 and 10 mag. Consequently, we delineate the interstellar visual extinction (A$_V$) range as 1-10 mag for SED fitting, accounting for uncertainties in our analysis.

Given the extensive number of SED models encompassing a broad parameter space, the fitting process for each source necessitates an increased number of data points and sufficient coverage across the entire wavelength range. Therefore, we selected sources with photometry available in at least all four IRAC bands as a criterion for SED modeling. It is important to note that photometry from \emph{Herschel} images was not utilized due to the absence of YSO counterparts at those wavelengths, primarily due to poor resolution. 

In addition to providing the best-fitting model, the SED fitting tool generates a set of well-fitted models ranked based on their $\chi^2$  values, which serve as a measure of their relative goodness-of-fit. Following a similar approach as \cite{robitaille2007interpreting}, we considered only those models that met the following criterion for further analysis:
\begin{equation}\label{robchi}
\chi^2  - \chi^{2}_{\rm min} < 3~({\rm per~data~point}).
\end{equation}
As described in \cite{robitaille2007interpreting}, it is worth noting that the criterion used for model selection, based on visual examination of SED plots, lacks rigid mathematical justification. Implementing a more stringent criterion could potentially lead to over-interpretation of the results. To determine the physical parameters, a weighted mean value and standard deviation were calculated for each parameter using the subset of models that satisfied equation (1). The inverse of the corresponding $\chi^2$ value was utilized as the weight for each model, following a similar approach as \cite{grave2009spitzer}. 



\subsubsection{Mass and Age spectrum}
\begin{figure}
\includegraphics[width=0.43\textwidth, trim= 0 0.1cm 0 0cm]{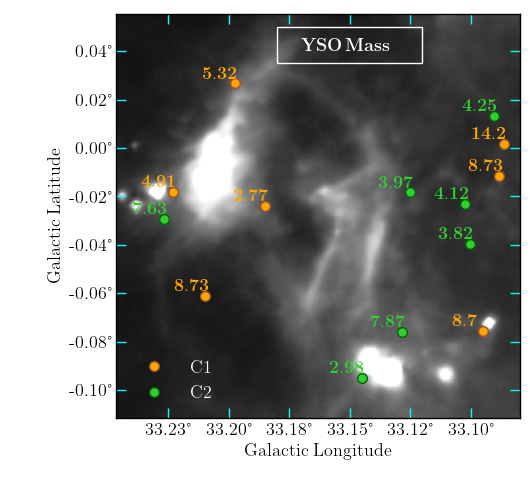} 
\includegraphics[width=0.43\textwidth, trim= 0 0.1cm 0 0]{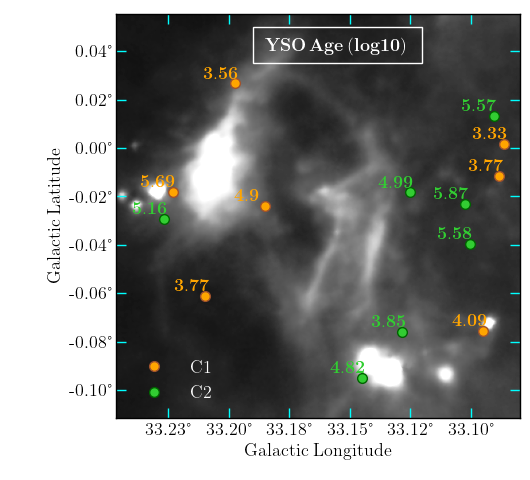}

\caption{The mass and age (in units of M$_\odot$ and year, respectively) of the YSOs in region R1 overlaid on PACS 70 \textmu m image, are displayed in the top and bottom panels. The age is displayed in the log10 scale.  Class I and Class II sources are marked in orange and green, respectively.} \label{fig:MassndAge}

\end{figure}

While we conducted SED fitting to all the recognized YSOs within the region R1, a subset of only 14 YSOs satisfied the $\chi^2$ criterion as defined by Equation~\ref{robchi}. This subset encompasses 7 Class I and 7 Class II sources. The visual representation in Fig.~\ref{fig:MassndAge} depicts these YSOs, displaying their respective mass and age properties, overlaid onto the PACS 70~\textmu m image. 
The orange and green filled circles represent Class I and Class II sources, respectively. The main parameters obtained from the SED analysis are given in Table~\ref{table_SEDparams}. The clump masses of these sources range from 2.77 - 14.20 M$_\odot$, and their ages span from 2.13 $\times$10$^4$ $-$ 7.46 $\times$10$^5$ years. Notably, four of our Class I sources exhibit high mass values, measuring at 14.20, 8.74, and 8.74 and 8.70~M$_\odot$ respectively. It is worth mentioning that the majority of YSOs observed in region R1 are characterized by relatively young ages, all of them falling below 1 Myr.
\begin{figure}
\includegraphics[width=0.47\textwidth, trim= 0 0.1cm 0 0cm]{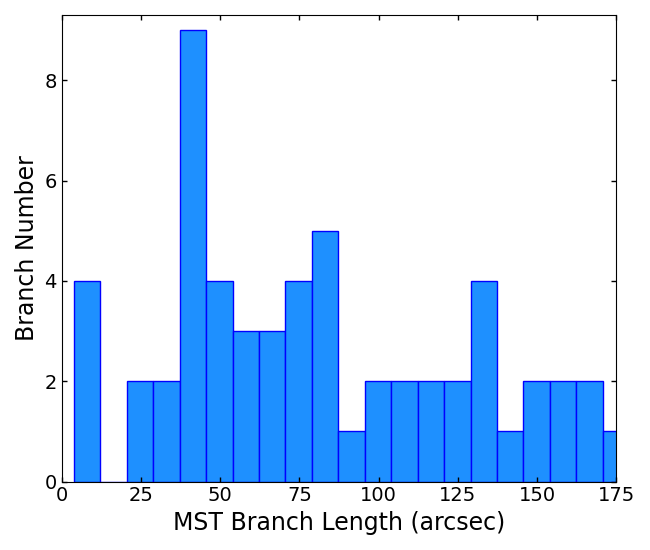} 
\caption{Histogram of the MST branch lengths used for critical length analyses of the YSOs.}
\label{fig:mstbranchist}
\end{figure}
\begin{figure*}
\begin{center}
\includegraphics[width=0.44\textwidth, trim= 0 0.1cm 0 0]{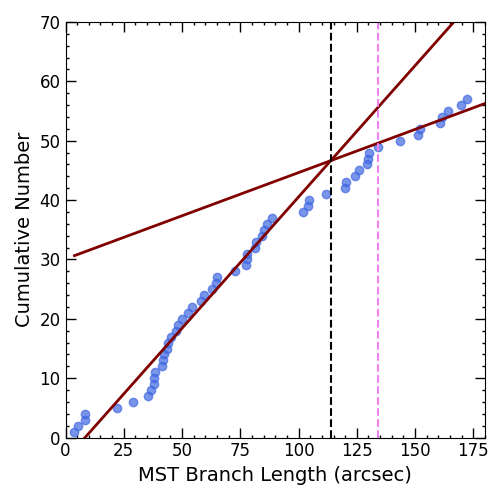}\quad
\includegraphics[width=0.47\textwidth, trim= 0 0.1cm 0 0]{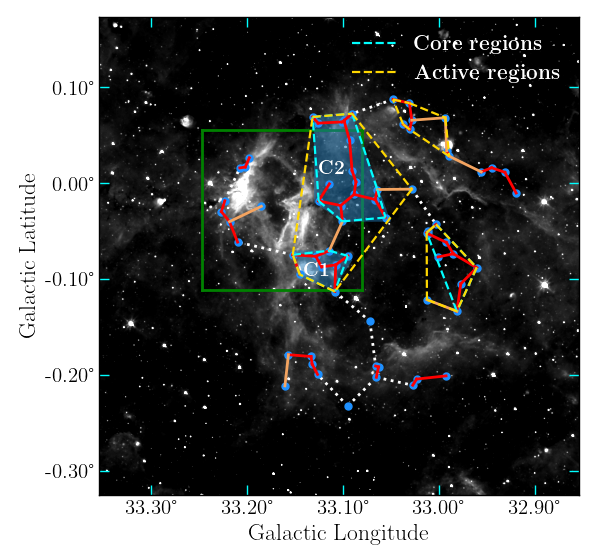}
\caption{Cluster identification using MST. The left panel shows the total number 
of sources that are connected by branches with lengths, $l$, shorter than the value on the $x$-axis. The steeper brown line is the linear fit to the data below 90\arcsec, and the other brown line is a similar fit to the data beyond 134\arcsec (vertical dashed-pink line). The cut-off length ($l_{\rm c}$) is opted as the branch length at the intersection of these lines (vertical dashed-black line). The image in the right panel shows the MST overlaid on \emph{Spitzer} 8.0~\textmu m map. The stars -- marked in blue -- are connected by solid red ($l<l_{\rm c}$), solid light-brown ($l_{\rm c}<l<134\arcsec$) and dotted 
white curves ($l>134\arcsec$), that together form the branches of the tree. The dashed-yellow and dashed-cyan polygons are convex hulls demarcating the active and core regions, respectively. The two core regions are shaded in blue and are labelled as C1 and C2. The green square shows the region R1.} \label{fig:mst}
\end{center}
\end{figure*}

\subsection{Clustering analysis}\label{cluster}

To investigate the star formation history of R1, it is imperative to identify distinct clusters or groups of YSOs. To accomplish this and assess the spatial distribution of YSOs, we employ the Minimum Spanning Tree (MST) method, which is predicated on the idea that
every point data set can be described by a unique, filamentary network \citep{Prim1957,Barrow1985}. It involves connecting the sources with branches such that their cumulative length is minimised, and the resultant tree is devoid of loops. Then, a cluster is defined as any subset of this network with (at least $N$) 
members whose associated branches are shorter than a specified cutoff length ($l_{\rm c}$). This enables cluster identification based solely on the spatial distribution of sources, without relying on kinematic information.

It has to be noted that MST is particularly effective with a larger number of data points. Therefore, we initially applied the method to the entire bubble N59, for a sample of 59~YSOs (the YSOs obtained in Section~\ref{S3} through (i) and (ii)) and subsequently narrowed our focus to examine the clustering patterns specifically within region R1. While there is not a strict minimum requirement for data points in constructing an MST, its significance lies in how well the data is interconnected. Notably, MST has been successfully applied to YSO samples, even with less than 50 data points (e.g., \citealt{Gutermuth2009,sharma2016structural,Verma_2023}). To gauge the reliability of the MST obtained, we examined the histogram illustrating the relationship between MST branch lengths and MST branch numbers for the YSOs (refer to Fig.~\ref{fig:mstbranchist}). The histogram reveals a peak at small spacings and a relatively long tail toward large spacings. Such peaked distance distributions often suggest significant subregions above a relatively uniform, elevated surface density. By implementing an MST length threshold (cutoff length, $l_{\rm c}$), we can isolate sources closer than this threshold.

Though this method has been successfully applied to numerous nearby star-forming regions \citep{Schmeja2006,Allison2009,Parker2012,Saral2015,Beuret2017}, we note that there is no universally prescribed criterion for determining the appropriate $N$ and $l_{\rm c}$. Nevertheless, it is common practice to inform $l_{\rm c}$ based on clustering of branch lengths \citep[e.g.][]{Koenig2008,Gutermuth2009,Saral2015,Pandey2020}, and we adopt the same for our analysis. We remind the reader that the cut-off length is not an extremely rigid value and may vary slightly as the data statistics increase. However, the key takeaway from this analysis remains the identification of a noticeable level of clustering in our specified region R1.

To this end, we first generate the cumulative distribution of the branch length, $l$. The left panel of Fig.~\ref{fig:mst} shows that the distribution comprises two quasi-linear zones separated by an intermediate zone. The
two quasi-linear zones span $0\arcsec<l<90\arcsec$ and $134\arcsec<l<175\arcsec$. The quasi-linear zone at smaller lengths and the intermediate zone essentially represent two levels of clustering, one at a smaller scale than the other. We derive linear fits to these zones (solid brown lines) and take the length corresponding to their intersection as the cut-off  -- i.e. $l_{\rm c}=114\arcsec$, which corresponds to 2.59~pc for a distance of 4.66 kpc.

Next, we use this cut-off to identify the small-scale clusters in the tree\footnote{This is computed using the \textsc{\large python} package at \url{https://github.com/jakevdp/mst_clustering}.} shown in the right panel of Fig.~\ref{fig:mst}, where the YSOs (in blue) are overlaid on the NIR surface density map of the region. The solid and dotted lines are the branches, where the red lines have lengths $l<l_{\rm c}$. We follow \citet{Saral2015} and use $N=7$, considering that this threshold is appropriate for avoiding false identifications stemming from random associations. Thus, the small-scale clusters are those overdensities in the red network that satisfy this threshold. These are demarcated using the dashed-cyan convex-hulls\footnote{We derive these using the \textsc{\large pyhull} package (\url{https://github.com/materialsvirtuallab/pyhull}).}(based on the Qhull algorithm; \citealt{Barber1996}). We term these as the `core' regions. 

Recall that the cumulative $l$ distribution suggests two levels of clustering. This is because the core regions are, in fact, substructures embedded in even larger regions that manifest as the intermediate zone in the left panel of Fig.~\ref{fig:mst}. These regions are traced by the brown branches in the right panel of Fig.~\ref{fig:mst}, corresponding to $l<$ 134\arcsec -- the lower bound of the quasi-linear zone at the
higher lengths (left panel; Fig.~\ref{fig:mst}). We define these `active' regions as overdensities in the brown network (again assuming $N=7$), depicted as dashed-yellow convex hulls. 

As illustrated in the right panel of Fig.~\ref{fig:mst}, we identify two core regions, labeled as C1 and C2, within R1. Core C1 is found to coincide precisely with a prominent dust condensation within the region, highlighting its significance in the context of star formation. Dust, a key player in the star formation process, fulfills a pivotal role by safeguarding molecular clouds from the harmful effects of ultraviolet (UV) radiation while facilitating the essential cooling mechanisms \citep{williams2005gas}. 
Remarkably, three out of the four massive YSOs identified through SED analysis are found within C1 and C2, suggesting that these regions are highly conducive to vigorous star formation. This is further buttressed by the presence of massive ATLASGAL clumps within these "cores". Furthermore, core C1 stands out with the presence of a 6.7 GHz methanol maser, further reinforcing the notion of active star formation underway within these cores. A more detailed discussion regarding the significance of ATLASGAL clumps and 6.7~GHz masers can be found in Section~\ref{impact}. 

\subsection{Ionized Emission}

\subsubsection{H\,II Regions}
\label{section_HIIregions}

\begin{figure*}
\centering
\includegraphics[width=0.9\textwidth]{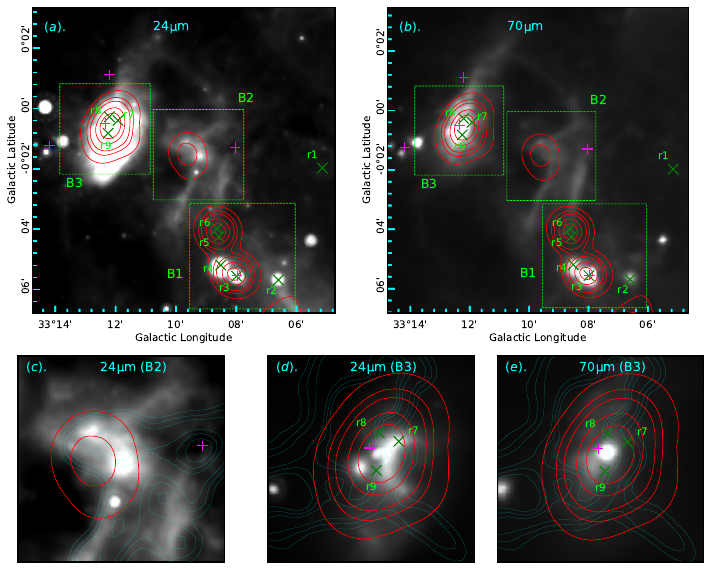}
\caption{
\emph{(a). Spitzer} 24~\textmu m, and (b). \emph{Herschel} 70~\textmu m images
with overlaid NVSS contours in red (at 0.01, 0.03, 0.05, 0.1, and 0.15 Jy/beam),
CORNISH and/or MAGPIS radio sources (crosses), and
ATLASGAL sources (plus symbols).
Three boxes, labelled B1, B2, and B3 mark three regions with NVSS emission.
(c), (d), and (e) show the zoomed-in view of the regions enclosed in boxes,
with the wavelength and box region labelled on the image.
Cyan contours show $\mathrm{^{13}}$CO integrated intensity emission at 1, 2, 3.5, 8, 10, and 
12 K\kms (see Section \ref{section_MolecularMorphology} and 
Fig. \ref{fig_MolecularSpectra} for details).} 

\label{fig_RadioSources}
\end{figure*}

\begin{figure}
\centering
\includegraphics[width=\columnwidth]{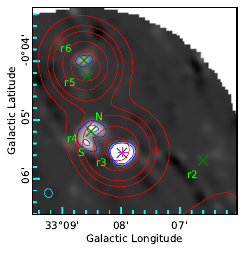}
\caption{
VLA 8.49\,GHz image of the B1 box region in Fig. \ref{fig_RadioSources}.
The symbols and red contours are same as in Fig. \ref{fig_RadioSources}.
Blue contours are VLA 8.49\,GHz image contours at 0.005 and 0.015 Jy/beam.
The radio beam is shown in bottom left corner. Here we can see the r4
source resolved into N and S components.
}
\label{fig_Radio8p49GHz}
\end{figure}

Fig. \ref{fig_RadioSources} shows the compact H\,II regions in the region from
NVSS (red contours), CORNISH, and MAGPIS. Sources from CORNISH and/or MAGPIS
have been marked in green crosses and labelled r1-r9.
Table \ref{table_RadioSources} shows the details of their detection.
In Figs \ref{fig_RadioSources}(a) and \ref{fig_RadioSources}(b) -- which
show the 24\micron\, and 70\micron\, emission from \emph{Spitzer} and
\emph{Herschel}, respectively -- the contours for NVSS 1.4\,GHz
emission (red contours) indicate three distinct H\,II regions, enclosed in
dashed boxes labelled B1, B2, and B3.
A zoomed-in view of the regions B2 and B3 is shown in
Figs \ref{fig_RadioSources}(c), \ref{fig_RadioSources}(d), and
\ref{fig_RadioSources}(e) for a better delineation of the features.
For the region B1, VLA archival image was
available at 8.49\,GHz, and is shown in Fig. \ref{fig_Radio8p49GHz}.
Emission at 24\micron\, and 70\micron\, is mainly due to heated dust,
such as that by the star formation
\citep{Calzetti_2013seg, Helfand_magpis_2006AJ}, and is thus helpful in
separating thermal (like H\,II regions) from nonthermal (like extragalactic
origin) sources. We now discuss the H\,II regions B1-B3 in detail.

For region B1, it can be seen that while the NVSS contours encompassing the
\textquotedblleft r3-r4\textquotedblright\, sources have associated thermal
dust emission at 24\micron\, and 70\micron, there is no such emission at
the location of contours associated with
\textquotedblleft r5-r6\textquotedblright\, sources.
Combined with the fact that the sources r5 and r6 have been marked as
Radio Galaxies in CORNISH catalogue (see Table \ref{table_RadioSources}),
it can be safely assumed that they are not associated with the N59 region.
While the NVSS contours (due to low resolution) subsume both r3 and r4 sources,
the background images clearly indicate them to be two separate sources.
High-resolution 8.49\,GHz image from VLA (Fig. \ref{fig_Radio8p49GHz})
also shows r3 and r4 to be distinct H\,II regions. Furthermore, in higher
resolution 8.49\,GHz image, the source r4 appears
to be resolved into two distinct compact H\,II regions
(beamsize here is $\sim$ 9\arcsec), labelled r4-N and r4-S in
Fig. \ref{fig_Radio8p49GHz}. The integrated flux density for r3, r4-N,
and r4-S was calculated using the AIPS software (task ``JMFIT''),
and values returned were used to calculate their spectral types
(see Section \ref{section_RadioSpectralTypes}).

Moving further north, we find weak 1.4\,GHz radio continuum emission in the
region marked by box B2.
It was found to be relatively faint and there are no CORNISH/MAGPIS
sources associated with this region.
The contours seem to be bounded by 24/70\micron\, emission towards their
western side (see the zoomed-in view in Fig. \ref{fig_RadioSources}(c)).
This could indicate some embedded source in the west, the
ionizing radiation from which is expanding as a possible blister/champagne flow
towards the east.

In Figs~\ref{fig_RadioSources}(a) and \ref{fig_RadioSources}(b), contours
in B3 region are associated with the brightest thermal dust emission.
Three MAGPIS 6\,cm sources were found to be present near the peak of the
1.4\,GHz emission, named r7, r8, and r9 here. The zoomed-in view in Figs~\ref{fig_RadioSources}(d) and \ref{fig_RadioSources}(e) does show a bright
source at the location of the peak, but it is not coincident with any of the
MAGPIS sources. Coupled with the fact that the r7-r9 have no detections
in MAGPIS 20~cm catalog and CORNISH catalog, it is probable that they
are spurious peaks.

Finally, we note the existence of two more CORNISH/MAGPIS sources, named r1
and r2 in Fig. \ref{fig_RadioSources}. Both these sources are only present
in MAGPIS 6\,cm catalog with no counterparts in either MAGPIS 20~cm or
CORNISH catalog. Both the sources have a high probability of being a sidelobe
according to the catalog.
In the light of these facts, the source r1 might merit a conclusion of being
spurious,
but for r2, we find that it has a corresponding source at both 24 and
70\micron.

\begin{table*}
\footnotesize
\centering
\caption{H\,II Regions}
\begin{tabular}{lcccccr}
\hline
Region & l       & b       & \multicolumn{2}{c}{MAGPIS}    & CORNISH & Remarks \\
       & (deg)   & (deg)   & 20\,cm      &  6\,cm          &         &         \\
\hline
r1     & 33.08575 & -0.03288  & -           &  $>$5.5$\sigma$ & -       & -        \\
r2     & 33.11007 & -0.09479 & -           &  $>$5.5$\sigma$ & -       &  B1-ZAMS \\
       &           &         &             &                 &          & using MAGPIS 6\,cm Flux \\
\hline
\multicolumn{1}{c}{B1} \\
\cmidrule(lr){1-1}
r3     & 33.13311  & -0.09270 & $>$5.5$\sigma$ & $>$5.5$\sigma$ & Highly Reliable H\,II Region & O9.5-ZAMS using 8.49\,GHz flux \\
       &           &          &                &                &                              & (Integrated Flux : 470.9$\pm$0.3\,mJy) \\
r4     & 33.14188  & -0.08655 & $<$5.5$\sigma$ & $>$5.5$\sigma$ & Highly Reliable H\,II Region & Resolved into 2 sources in 8.49\,GHz image \\
\cmidrule(lr{2.75em}){1-1}
~~~r4-N & 33.14138  & -0.08615 & -              & -              & -                            & B0.5-ZAMS using 8.49\,GHz flux \\
        &           &          &                &                &                              & (Integrated Flux : 57.9$\pm$0.6\,mJy) \\
~~~r4-S & 33.14323  & -0.08942 & -              & -              & -                            & B0.5-ZAMS using 8.49\,GHz flux \\
        &           &          &                &                &                              & (Integrated Flux : 47.1$\pm$0.5\,mJy) \\
r5     & 33.14297  & -0.07085 & $>$5.5$\sigma$ & -              & Highly Reliable Radio Galaxy & - \\
r6     & 33.14384  & -0.06639 & $>$5.5$\sigma$ & $>$5.5$\sigma$ & Highly Reliable Radio Galaxy & - \\
\hline
\multicolumn{1}{c}{B2} \\
\cmidrule(lr){1-1}
NVSS Source    & 33.16007 & -0.02546 & -         & -              & -       & B0.5-ZAMS \\
\hline
\multicolumn{1}{c}{B3} \\
\cmidrule(lr){1-1}
NVSS Source  & 33.20282  & -0.00978 & -         & -               & -       & B0-ZAMS \\
r7           & 33.19871  & -0.00658 & -         &  $>$5.5$\sigma$ & -       & -       \\
r8           & 33.20343  & -0.00452 & -         &  $>$5.5$\sigma$ & -       & -       \\
r9           & 33.20416  & -0.01366 & -         &  $>$5.5$\sigma$ & -       & -       \\
\hline
\end{tabular}
\label{table_RadioSources}
\end{table*}

\subsubsection{Spectral Types}
\label{section_RadioSpectralTypes}

In this section, we calculate the Lyman continuum luminosity for the sources
likely to be H\,II regions.
Apropos the discussion in Section \ref{section_HIIregions}, this includes
r2, r3, r4-N, r4-S, and NVSS sources in regions B2 and B3.
The following expression from \citet{Moran_Radio_83} was used for the same :
\begin{equation}
N_{c} = 8 \times 10^{46} \left(\frac{S}{Jy}\right) \left(\frac{D}{kpc}\right)^2 \left(\frac{\nu}{GHz}\right)^{0.1} \left(\frac{T_e}{10^4 K}\right)^{-0.45}
\end{equation}
where
N$_c$ is the Lyman continuum luminosity in photons s$^{-1}$, 
S is the flux density in Jy, D is the distance in kpc, $\nu$ is the frequency
in GHz, and T$_e$ is the electron temperature.
While for r2, we used the MAGPIS 6\,cm (integrated) flux for the above
calculation; for r3, r4-N, and r4-S, we use the 8.49\,GHz integrated flux
obtained using the AIPS software (``JMFIT'' task). For the NVSS sources in
B2 and B3 boxes, we use the integrated flux from the NVSS catalog.
The sources r2, r3, r4-N, r4-S, B2 NVSS source, and B3 NVSS source were
found to be of spectral types B1, O9.5, B0.5, B0.5, B0,5, and B0, on a
comparison of N$_\mathrm{c}$ with the tabulated values of
\citet[assuming ZAMS]{Panagia_73}.
The results are given in Table \ref{table_RadioSources}. Hence there appears
to be a rich cluster of O and B type stars in the region.


\subsection{Molecular Cloud Morphology}
\label{section_MolecularMorphology}

\begin{figure*}
\centering
\includegraphics[width=0.9\textwidth]{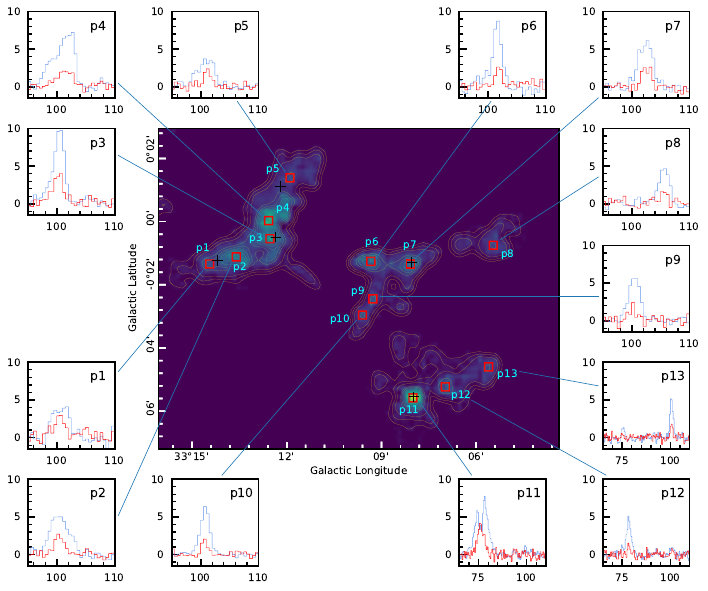}
\caption{
Integrated intensity \tco\, map of the region, showing only those regions
with emission above 3$\sigma$ threshold. The (smoothed) image contours have
been drawn at 1, 2, 3.5, 8, 10, 12, 14, 17, 25, 30, and 40 K\kms.
Local integrated intensity peaks have been marked by boxes and labelled
p1-p13. \tco\, (blue) and \cetno\, (red) spectra at each of these peak
locations has been extracted. Plus symbols mark the ATLASGAL clumps.
}
\label{fig_MolecularSpectra}
\end{figure*}

\begin{figure*}
\centering
\includegraphics[width=0.8\textwidth]{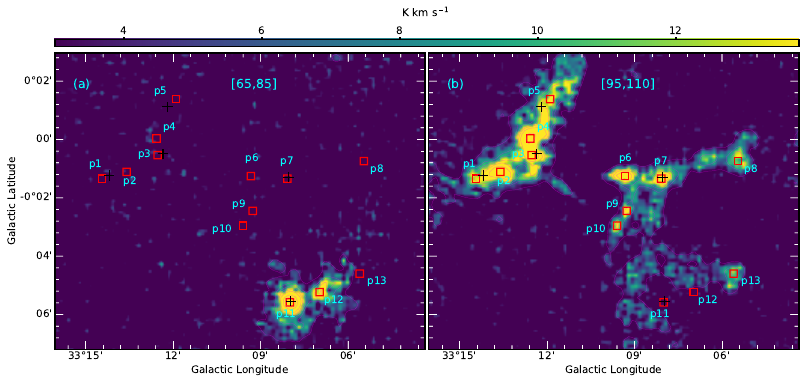}
\caption{
Channel maps for the velocity ranges encompassing
the two velocity components in spectra
(Fig. \ref{fig_MolecularSpectra}). The (smoothed) contours are
shown at 3.5, 8, 10, 12, 14, 17, 25, 30, 40, and 45 K\kms.
The symbols are same as Fig. \ref{fig_MolecularSpectra}.
}
\label{fig_13CO_ChannelMaps}
\end{figure*}

\begin{figure*}
\centering
\includegraphics[width=1.0\textwidth]{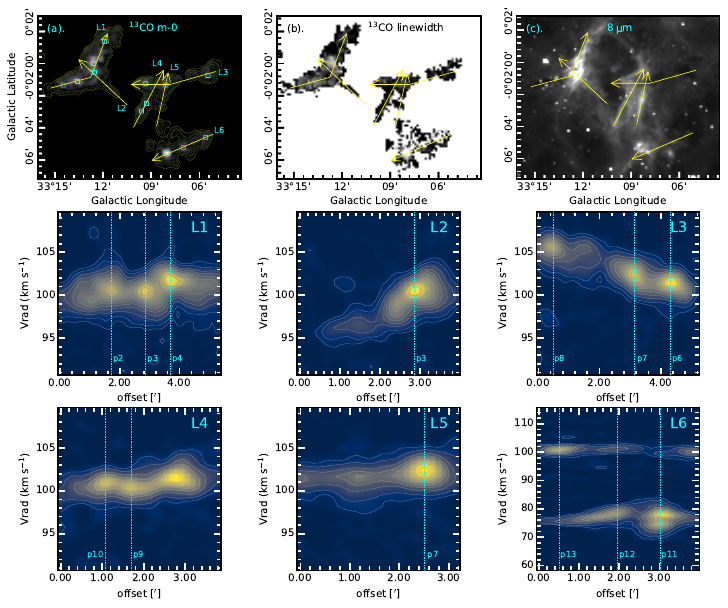}
\caption{
\emph{(a)} Integrated intensity \tco\, map of the region from
Fig. \ref{fig_MolecularSpectra}, with (smoothed) image contours
drawn at 1, 2, 3.5, 8, and 10 K\kms.
Cyan boxes are the local integrated intensity peaks as in Fig.
\ref{fig_MolecularSpectra}.
Figs
\emph{(b)} and \emph{(c)} show the \tco\, linewidth and \emph{Spitzer}
8\micron\, image of the region, respectively, for comparison.
The green vectors, named L1-L6, mark the regions along which \tco\,
position-velocity slices were extracted. These slices are shown in
bottom two rows. Contours (smoothed) have been drawn at
0.25, 0.5, 1, 1.5, 2, 2.5, 3, 3.5, 4, and 5 K for these pv slices.
Vertical dashed lines (labelled) mark the locations of integrated 
intensity peaks (also see Fig. \ref{fig_MolecularSpectra}).
}
\label{fig_PVmaps}
\end{figure*}

We use the J=3-2 transition of $^{13}$CO and C$^{18}$O isotopologues to examine
the molecular structure in the region. This transition has a critical density
$>$\,10$^4$cm$^{-3}$ \citep{Rigby_CHIMPS_2016MNRAS}, with \cetno\, tracing
the densest structures in the region.
We first examine the integrated intensity map (moment-0) of the region
in \tco\, transition in Fig. \ref{fig_MolecularSpectra}.
To avoid confusing artefacts with real features, this image is made using
a masked cube of only those regions which show detection above 3$\sigma$
($\sigma$ being the noise level of the cube) limit in the
position-position-velocity (ppv) space. This was achieved by identifying
clumps in the ppv space above the requisite threshold -- using the
clumpfind algorithm -- and then masking the cube to only show those regions.
The masked cube 
was collapsed in the velocity range $\sim$\,[65, 110]\kms\, (as the molecular 
emission was found to be in this range only for our region of interest) 
to make the moment-0 map shown in Fig. \ref{fig_MolecularSpectra}.
We also obtained the linewidth image from the masked cube in this velocity 
range (discussed in Section \ref{section_PVMaps}).

The image has been overlaid with contours showing local peaks (p1-p13),
with spectra extracted therein in \tco\, (blue) and \cetno\, (red)
transitions. ATLASGAL sources have also been marked in plus symbols,
and seem to be associated with $\sim$p1/p2, p3, $\sim$p5, p7, and p11.
We examine the \tco\, spectra now.
The first thing to notice in the extracted spectra is that many of them
have complex shapes unlike what a simple gaussian would indicate.
For example, while p1, p2, p5, and p8 have broad peaks; p3 and p4 have wide
wings towards the blue-shifted velocities; and p11 shows a sharp
self-absorbtion feature near the peak.
The velocity peaks at p11, and p12 positions show a marked contrast
to other regions. While the main velocity peak for p1-p10 regions lies at
$\sim$95-110\kms, the same for p11 and p12 lies at $\sim$70-80\kms.
Among the regions which have velocity peak in $\sim$95-110\kms\, range, 
the peaks p4 and p5 have slightly red-shifted velocities as compared to 
p1, p2, and p3; with similar trend for p6, p7, and p8 as compared to 
p9 and p10 (i.e. slightly red-shifted velocity). 

It can be noticed that while p1, p2, p3, p9,
and p10 have peaks in the range $\sim$98-102\kms; p4, p5, p6, and p7 have
the peak in $\sim$100-104\kms range; and p8 has the same in $\sim$104-108\kms\,
range.
At p13, both the velocity components can be seen, albeit the $\sim$100\kms\,
the component is much stronger.
This suggests two different molecular clouds in the region, which spatially
might appear as part of the same complex, but are differentiated in
velocity space.

To examine the morphology of the two velocity components, we constructed
channel maps (using the native cube) for the velocity ranges [65,85]\kms\,
and [95,110]\kms\, (Fig. \ref{fig_13CO_ChannelMaps}). The channel maps
clearly reveal two molecular cloud regions. It is worthwhile to note that
there is negligible overlap between the spatial locations of the two
molecular cloud structures.
In fact, in Fig. \ref{fig_13CO_ChannelMaps}(b), there is a noticeable
``hole'' at the location of the blue-shifted velocity cloud.
The \tco\, integrated intensity peaks p11 and p12 are associated with the
blue-shifted molecular cloud.

\subsubsection{Position-Velocity Maps}
\label{section_PVMaps}

The \tco\, position-velocity (pv) maps along different
-- seeemingly continguous --
structures have been shown in Fig. \ref{fig_PVmaps}.
The pv slices have been extracted along six different cuts, named
L1-L6 in Fig. \ref{fig_PVmaps}(a) (\tco\, integrated intensity
map from Fig. \ref{fig_MolecularSpectra}). For comparison, the
\tco\, linewidth map and the \emph{Spitzer} 8\micron\, emission map
for the region have been shown in Figs \ref{fig_PVmaps}(b) and
\ref{fig_PVmaps}(c), respectively.

Here we notice that while the structures are contiguous in position-velocity (pv)
space, there is considerable fluctuation along the velocity axis for most
of the pv maps, as one traverses along the line.
These fluctuations are most prominent along the directions L1, L2, and L3.
For L1, while there is spread in velocity along both sides of $\sim$100\kms\,
velocity; for L2 and L3, the contiguous structure is blueshifted and
redshifted, respectively.
Normally, such a shape would imply velocity gradients, and thus a longitudinal
flow of matter along the structure. Along L2, the clump at $\sim$2.6\arcmin\,
corresponds to p3 position (also see Fig. \ref{fig_MolecularSpectra}),
which is also associated with an ATLASGAL clump.
Along L3, there is clumping at $\sim$0.4\arcmin, 3.2\arcmin, and 4.4\arcmin,
associated with positions p8, p7 (also an ATLASGAL clump), and p6.
Since in the linewidth map (Fig. \ref{fig_PVmaps}(b)), there is
higher velocity dispersion at p3 (along L2), thus it is likely that the
flow of matter is along the L2 molecular filament towards the gravitational
well of p3.
Similarly for L3, the relatively higher velocity dispersion at p7 and p6
would suggest longitudinal flow along the direction of the vector, though
only p7 is associated with an ATLASGAL clump here.
The structures along the vectors L4 and L5 appear to be elongated filamentary
structures, and seem to be joining L3 near peaks p6 and p7.
The gradient along L4 and L5 is much less than the case for L2 and L3.
While the \tco\, molecular emission along L5 seems fragmented, it appears
as a prominent contiguous structure in \emph{Spitzer} 8\micron\, emission,
as well as at 24 and 70\micron\, (see Fig. \ref{fig_RadioSources}).
Finally, for L6, the pv slice shows two distinct clouds, as was also
discussed earlier (see Fig. \ref{fig_13CO_ChannelMaps}). The redshifted
cloud ($\sim$100\kms) appears to have a uniform velocity throughout, while
the blueshifted cloud has clumps associated with positions p12
(at $\sim$2\arcmin) and p11 (at $\sim$3\arcmin). There is a small velocity
gradient up to p12, and p11 shows a double-peaked structure indicating
self-absorbtion, as was also seen in the spectrum (see Fig.
\ref{fig_MolecularSpectra}). p11 also has an associated ATLASGAL clump
and shows a large dispersion in Fig. \ref{fig_PVmaps}(b).

\section{Discussion}\label{dis}
\label{section_Discussion}
\subsection{Implications and Insights from 6.7 GHz masers}\label{impact}
\begin{figure}
\includegraphics[width=0.45\textwidth, trim= 0 0.1cm 0 0cm]{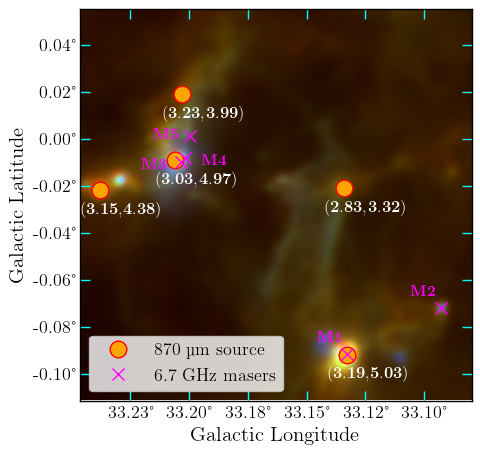} 
\caption{870 \textmu m ATLASGAL clumps (orange circles) and locations of 6.7 GHz maser emissions (magenta crosses) overlaid on \emph{Herschel} three colour composite map (250~\textmu m (red), 160~\textmu m (green), and 70~\textmu m (blue) images in linear scale)).
The value pairs are log10 masses and lumnosities in units of M$_\odot$ and L$_\odot$.}\label{fig:Maserspots}
\end{figure}
\begin{figure}
\includegraphics[width=0.47\textwidth, trim= 0 0.1cm 0 0cm]{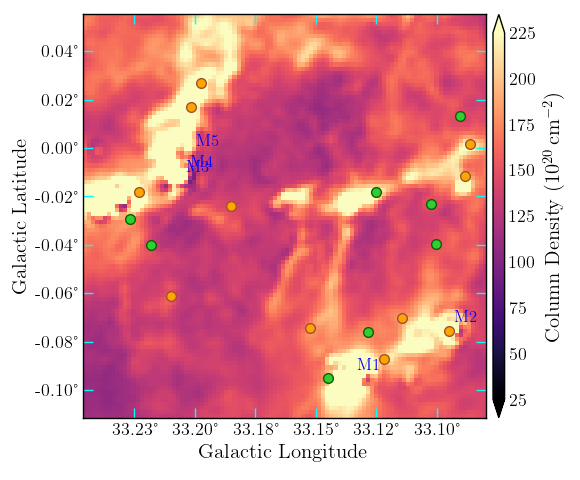} 
\includegraphics[width=0.47\textwidth, trim= 0.1cm 0.1cm 0 0]{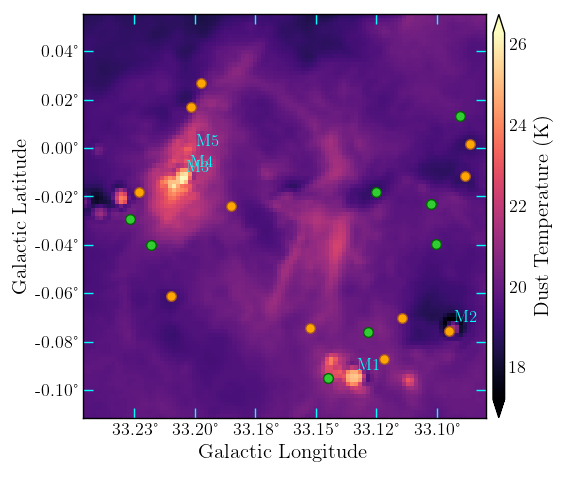}
\caption{Top and bottom panels illustrate column density and temperature maps for region R1. The column density map is in the unit of 10$^{20}$ cm$^{-2}$. Class~I and Class~II YSOs are marked by orange and green filled circles, respectively, while labels M1-M5 indicate the positions of 6.7 GHz methanol masers.} \label{fig:colmaps}
\end{figure}
According to \citet{deharveng2010gallery}, bubble N59 emerges as a faint structure spanning a diameter of approximately 20 parsecs. Our observations highlight a distinct concentration of star-forming activity primarily situated in the north eastern region of the bubble, marked as R1.  Worth highlighting is the presence of five 6.7 GHz methanol masers within R1's bounds (refer to Fig.\ref{fig:Maserspots}). Intriguingly, two of these maser detections align precisely with the 870~\textmu m dust continuum clumps, exhibiting luminosities exceeding 10$^4$ L$_\odot$. This discovery is consistent with the conclusions drawn by \citet{bourke2005identification}, suggesting a lower luminosity threshold of approximately 10$^3$ L$_\odot$ for a source to host 6.7~GHz methanol masers. Our photometric analysis identifies a massive Class~I YSO (M$\sim$8.7~M$_\odot$; hereafter HM1) in the proximity of maser M2. However, an angular offset of approximately 13\arcsec~between their positions raises questions about the precise association between the maser and the identified YSO. As previously mentioned, the presence of 6.7 GHz masers typically indicates the early stages of massive star formation. This inference is reinforced by the observed dust temperature values within the maser hosting regions. For example, Fig.~\ref{fig:colmaps} displays the maps illustrating the column density and temperature distribution of R1 (from ViaLactea project, see Section \ref{vialectea}).

Notably, the column density values within this region are found to be approximately on the order of 10$^{22}$ cm$^{-2}$. The top panel of Fig.~\ref{fig:colmaps} clearly depicts that the majority of YSOs are situated within areas characterized by significantly higher column densities (>10$^{22}$ cm$^{-2}$). Similar column density values exceeding 10$^{22}$ cm$^{-2}$ have been reported in numerous instances of massive star-forming regions, particularly during their very early stages of star formation (e.g. \citealt{rygl2010initial}, \citealt{gieser2021physical} and references therein).

It is noteworthy that while not all peaks in the column density map coincide with peaks in the temperature map, the locations of methanol masers are consistently associated with regions exhibiting both high column density and elevated temperature (with temperatures > 18~K; \citet{paulson2020probing}). Of particular interest is the column density and dust temperature observed towards HM1. In this context, the recorded values are approximately 3.6 $\times$ 10$^{22}$ cm$^{-2}$ and 23.7~K, respectively. Similar dust temperature and column density values have been reported towards other massive star forming sites \citep{yu2019multiwavelength,bally2010herschel}. Dust temperature serves as a dependable indicator of a source's evolutionary stage, with maser clouds generally displaying higher temperatures than less developed Infrared Dark Clouds but consistently lower temperatures than more evolved sources such as $\HII$~regions \citep{giannetti2013physical} In the case of M1 and M2, as derived from \emph{Herschel} dust temperature maps, the average dust temperature is approximately 21~K. This finding strongly suggests that these sites are positioned in the early stages of massive star formation, characterized by conditions conducive to the birth of massive stars \citep{paulson2020probing, urquhart2014atlasgal}.

Furthermore, it is noteworthy that the ATLASGAL clump masses exhibit a range from approximately 6$\times$10$^2$ to 10$^3$~M$_\odot$. The mean clump mass within this sample stands at approximately 1300~M$_\odot$, a sufficient mass to give rise to a star cluster that includes at least one massive star with a stellar mass of $\geq $10 M$_\odot$ \citep{urquhart2014atlasgal}. Particularly intriguing is the affiliation of both maser sources, M1 and M2, with the core region C1 (Refer Fig.~\ref{fig:mst}), implying that this region may host massive stars or be actively involved in ongoing massive star formation processes. Notably, targeted ALMA archival data towards methanol maser M1 exists, and the analysis of the data holds promise for revealing compelling underlying physical phenomena. 

\subsection{Star formation scenarios in R1} \label{starformationR1}
The brightest condensation of bubble N59 is positioned towards its northern edge, as depicted in box B3 of Fig.~\ref{fig_RadioSources}. 
In B2 (refer to Fig.~\ref{fig_RadioSources}(c)), there is a potential indication of a blister or champagne-like flow structure of the $\HII$ region. These structures commonly become evident when a massive star takes shape near the periphery of a molecular cloud. In such scenarios, the ionized gas either diffuses into the inter-cloud medium at the cloud's edge or follows a density gradient within the star's birth cloud. 

It is possible that the blue-shifted tail seen at p6 (see Fig. \ref{fig_MolecularSpectra} and the L3 pv map in Fig.~\ref{fig_PVmaps}) could indicate the molecular material pushed towards the observer by the 
champagne flow. 

The high-mass star formation in B3 (refer Fig.~\ref{fig_MolecularSpectra}) could be acting as a sink of material along vector L2. This is substantiated by the prominent velocity gradient observed along L2. Similarly, notable velocity gradients are also discernible in the vicinity of p6 and p7 along L3, L4, and L5. These gradients could suggest the directed flow of gas toward these sources. This phenomenon of gas flow along filamentary structures towards sources mirrors the longitudinal flow observed in filaments towards central hubs within the hub-filament system. The association of p7 with an ATLASGAL clump stands out, with its column density estimated to be approximately 10$^{22}$ cm$^{-2}$ \citep{urquhart2022atlasgal}. Such regions with comparable column densities have been classified as "hubs" within the context of hub-filament systems, as per existing literature \citep{myers2009filamentary, Kumar_2020AA}. Consequently, vectors L3, L4, and L5 seem to be integral components of a hub-filament system, wherein p6 and p7 are linked with hubs. This association implies that the region related to B2 could serve as an illustrative case for exploring diverse star formation models. However, delving deeper into understanding the molecular cloud evolution as it gives rise to stars requires further investigation. This involves employing high-resolution studies encompassing molecular transitions of different species, examinations of magnetic fields, and similar approaches to comprehensively grasp the intricate processes at play.

Particular interest is drawn to p11, highlighted in Fig.~\ref{fig_MolecularSpectra}, due to its distinctive attributes. Notably, this cloud exhibits a primary velocity peak at approximately 75 km/s, in stark contrast to the peak velocity of 100 km/s observed in regions p1-p10. An interesting characteristic is the red-asymmetric line profile evident in $^{13}$CO, which contrasts with the single-peaked profile detected in C$^{18}$O, specifically near the central dip of the $^{13}$CO profile. This "red profile," where the redshifted peak of a double-peaked profile is more pronounced for the optically thick line, indicates outflow motions, caused by the absorption of colder blueshifted gas in front of the hot core \citep[e.g.][and the references therein]{1998sao..rept.....M,liu2011infall}. In their study of the high-mass star-forming complex G9.62+0.19, \citet{liu2011infall} note that around younger cores, the outflow is more active and colder than the more developed ultra-compact $\HII$ (UC $\HII$) regions. This results in a more pronounced "red profile." In the more mature UC~$\HII$ regions, where outflows are weaker, the surrounding gas is thermalized, and the temperature gradient toward the central star is more likely to cause a “blue profile", leading to a higher blue excess compared to the early stages of UC~$\HII$ or UC~$\HII$ precursors. Considering that the 6.7~GHz methanol maser signals the hot molecular core phase \citep[and the very early stage of UC~$\HII$ region -- e.g.][]{walsh1998studies,minier2001vlbi,ellingsen2006methanol} of high-mass star formation, and given that site p11 hosts a 6.7 GHz methanol maser, the presence of the "red profile" at p11 is well-supported. 

Additionally, the most massive star pinpointed in R1, an O9.5 star, is situated in proximity to p11 and exhibits detectable 6.7 GHz maser emission nearby. This locale presents itself as a promising candidate for subsequent high-resolution observations aimed at probing the underlying mechanisms of massive star formation. Further scrutiny reveals that while p11 and p12 exhibit velocities around ~70-80 km/s, p13, situated in close proximity to these regions, displays a peak velocity of approximately 100 km/s. This disparity introduces an intriguing scenario of two distinct molecular clouds that, despite their spatial proximity, are differentiated in velocity space. While the possibility of star formation through cloud-cloud collapse might have been conceivable, the notable velocity difference exceeding ~15 km/s between the cloud velocities necessitates the dismissal of this scenario.

R1 presents another captivating element with the existence of a bright-rimmed cloud (BRC)-like structure, found towards the site p6 in B2 (refer Figs~\ref{fig_RadioSources}(c) and \ref{fig_MolecularSpectra}). What makes this feature intriguing is its configuration, where the "head" of the BRC is oriented towards the bright dust condensation (Refer bottom panel of Fig.~\ref{fig:colmaps}) located in the northern region of R1. Furthermore, B2 exhibits a blister-like morphology, with its ionized region expanding towards the east. This expansion suggests the potential presence of embedded massive star formation in the western part of B2, contributing to the formation of B2's $\HII$  region. It is plausible that this embedded star formation in B2 is triggered by the adjacent $\HII$ region of B3. This connection to ionized emission is significant because BRCs are known to originate from the influence of such ionization, often leading to triggered star formation \citep{choudhury2010triggered,fukuda2013triggered,panwar2020star}. Additionally, the detection of NVSS emission in B2 indicates the possibility of ongoing robust star formation, potentially indicating a scenario of radiation-driven implosion. Alternatively, these illuminated bright rims may serve as the delineations of regular-density filaments shaped by ionizing radiation, all converging towards the central area.

\subsection{General overview of star formation in and around bubble N59}
Finally, we want to discuss the overarching bubble morphology of bubble N59 and the associated star formation processes within its vicinity. It is noteworthy to highlight that a considerable number of YSOs identified in Section~\ref{S3} are located on the border of the bubble, as depicted in Fig.~\ref{fig:mst} (blue dots in right panel). 
These YSOs are likely influenced by the N59 bubble itself. Three prominent processes have been considered in the literature regarding star formation at the periphery of the $\HII$ regions: collect and collapse (CC; \citealt{elmegreen1977sequential}), radiation-driven implosion (RDI; \citealt{sandford1982radiation}), and cloud-cloud collision (CCC) (e.g.,~\citealt{habe1992gravitational};~\citealt{haworth2015isolating};~\citealt{torii2017triggered}). In the CC process, a compressed layer of high-density neutral material forms between the ionization front and shock front, leading to regularly spaced fragments along the molecular ring or shell. RDI involves shocks driving into pre-existing density structures, forming stars characterized by cometary globules or optically bright rims. CCC can result in the rapid accumulation of cloud mass into a small volume, forming massive star-forming clumps within the shock interface.

Bubble N59 exhibits a broken ring/shell-like structure of gas and dust surrounding the cluster of massive stars, with several YSOs at the edges, suggesting the possible impact of O and B type massive stars located in the R1 region or other areas within the bubble. Notably, \citet{hanaoka2019systematic} reported a total infrared (TIR) luminosity, log (L$\mathrm{_{TIR
}}$/L$_{\odot}$) $\sim$ 6.6  , which may not solely originate from the massive stars within the R1 region. This high luminosity could be attributed to a cluster of massive stars located towards the centre of bubble N59. Additionally, the overall morphology indicates that the expansion of an older bubble might have triggered the star formation, leading to the development of the more recent sub-region R1. This further supports the potential presence of an older massive star population at the center of bubble N59. Although, no such massive star clusters were identified at the center of the bubble.

However, it's crucial to note that the massive stars responsible for the development of the bubble may not necessarily lie at the center of the bubble. \citet{hanaoka2019systematic} proposed a scenario where, for bubbles in the inner Galactic region, the heating sources are positioned nearly opposite to the broken sector directions. This offset could be attributed to the easier expansion of $\HII$ regions toward the lower density direction, causing the lower density side of the shell to break. Another explanation by \citet{hattori2016mid} links the offset of heating sources to the formation of massive stars on the boundary of collided clouds, resulting in a broken structure. Interestingly, bubble N59 hosts two Galactic filaments \citep{li2016atlasgal} and a massive protostar near the broken edge (at Galactic coordinates {\ensuremath{l=32.996}$^\circ$; \ensuremath{b=0.042}$^\circ$}) as catalogued by the RMS\footnote{Red MSX Source (RMS) survey; \url{http://rms.leeds.ac.uk/cgi-bin/public/RMS_DATABASE.cgi}} survey (\citealt{2016yCat..74524029U}) at its edges. \citet{hanaoka2019systematic} suggest that if the CCC process is essential for broken bubbles in inner Galactic regions, it is expected that their massive stars would be substantially offset to the opposite side of the broken sector direction.

Further in-depth investigation is required to elucidate the exact mechanism by which bubble N59 formed, which is currently beyond the scope of this paper. In their statistical analysis of 65 IR bubbles, \citet{deharveng2010gallery} found that over a quarter of the bubbles triggered the formation of massive objects, either through collect and collapse or via the compression of pre-existing clouds (CCC). Therefore, star formation triggered by $\HII$ regions may play a significant role, particularly in the formation of massive stars.

\label{S7}

\section{Summary and Conclusion}\label{summary}
\label{section_SummaryConclusion}
Bubble N59 emerges as a remarkably rich and diverse site encompassing a multitude of 6.7 GHz methanol masers, ATLASGAL clumps, $\HII$ regions, filaments, and bright-rimmed clouds. Our comprehensive analysis, including SED analysis and temperature mapping, indicates that region R1 within bubble N59 represents a relatively young and active massive star-forming region. The important conclusions of this work are as follows:

\begin{enumerate}    
\item We infer the distance towards the bubble to be $4.66 \pm 0.70$~kpc, using the \emph{Gaia} DR3 data (Section~\ref{dist}).
\item The MST cluster analysis reveals the presence of two prominent core regions containing multiple $\HII$ regions and massive YSOs (Section~\ref{cluster}). 
\item We observe a coexistence of one O-type star and five B-type stars, along with numerous low mass stars and a minimum of four high-mass (M > 8 M$\odot$) YSOs, indicating a unique stellar population within the region (Section~\ref{section_RadioSpectralTypes}).
\item The presence of a bright-rimmed cloud, likely formed due to feedback from the intense $\HII$ region, further adds to the intriguing nature of the region. The NVSS emissions hint at the ongoing massive star formation inside the bright-rimmed cloud-- a possible case of radiation-driven implosion (Section~\ref{starformationR1}).
\item The area surrounding B2 presents a compelling scenario suggestive of a hub-filament system, with the \tco\, local peaks p6 and p7 acting as the central "hub." Velocity gradient was observed towards these "hubs" along three filamentary structures, resembling the longitudinal flow observed in filaments leading to central hubs within hub-filament systems. Additionally, p7 is associated with an ATLASGAL clump, exhibiting a comparable column density to that of "hubs." These findings make the B2 region an intriguing subject for investigating various star formation models (Section~\ref{starformationR1}).

\item We identify two spatially close molecular clouds, each characterized by distinct velocities of 70-80 km/s and 100 km/s (see Fig.~\ref{fig_13CO_ChannelMaps}). Additionally, source p11 is found to have a 6.7~GHz methanol maser detection within an angular distance of less than 14$\arcsec$ (Section~\ref{impact} \& \ref{starformationR1}). To gain a deeper understanding of the mechanisms driving massive star formation in this region, high-resolution molecular studies are crucial.
\end{enumerate}

It's important to highlight that while a portion of the region (towards maser M1) has been observed using facilities like ALMA, there is a need for comprehensive studies encompassing various molecular transitions, magnetic fields, and other relevant factors to unravel the complex processes involved in massive star formation throughout the entire region R1.

\section*{Acknowledgements}
We thank the anonymous referee for providing suggestions that have improved the quality of the paper. STP and DKO acknowledge the support of the Department of Atomic Energy, Government of India, under Project Identification No. RTI 4002. This research has  made  use of NASA's Astrophysics Data System and VIZIER service operated at CDS, Strasbourg, France. The \textsc{\textmd{matplotlib}} package \citep{Hunter2007} for \large python was used for making plots. We have used the online freemium academic writing environment Overleaf available at: https://www.overleaf.com/.

\section*{Data Availability}
 We obtained the archival reduced and calibrated JCMT (James Clerk Maxwell Telescope) spectral cubes associated with the CHIMPS survey from the Canadian Astronomy Data Centre (CADC) (\url{https://www.cadc-ccda.hia-iha.nrc-cnrc.gc.ca/en/}). The PACS and SPIRE data is available at the ESA Hershel Science Archive (\url{http://archives.esac.esa.int/hsa/whsa/}). The ATLASGAL data can be obtained at \url{https://atlasgal.mpifr-bonn.mpg.de/cgi-bin/ATLASGAL_DATASETS.cgi}. The UKIDSS images and catalogs are available through a fully queryable user interface - the WFCAM Science Archive
(WSA : \url{http://surveys.roe.ac.uk/wsa}). The \emph{Gaia} data can be retrieved from the archive website at \url{https://archives.esac.esa.int/gaia}. 

\bibliographystyle{mnras}
\bibliography{bib_n59.bib}
\appendix
\section{SQL CONDITIONS FOR UKIDSS SOURCE SELECTION} \label{sql}
SELECT sourceID, ra, dec, l, b, jmhPnt, jmhPntErr, hmk\_1Pnt, hmk\_1PntErr,mergedClass, pStar, jAperMag3, jAperMag3Err, jEll, jppErrBits, jXi, jEta, hAperMag3, hAperMag3Err, hEll, hppErrBits, hXi, hEta, k\_1AperMag3, k\_1AperMag3Err,k\_1Ell, k\_1ppErrBits, k\_1Xi, k\_1Eta
FROM gpsSource
WHERE l BETWEEN 32.9 AND 33.3
AND b BETWEEN -0.25 AND 0.15
AND mergedClass != 0
AND (PriOrSec=0 OR PriOrSec=framesetID)
AND pstar>0.99 
AND jppErrbits<256 
AND hppErrbits<256 
AND k\_1ppErrbits <256 
AND sqrt(hXi*hXi + hEta*hEta)<0.3 
AND sqrt(k\_1Xi*k\_1Xi + k\_1Eta*k\_1Eta)<0.3
AND jEll<0.2 
AND hEll<0.2 
AND k\_1Ell<0.2
AND jAperMag3Err<=0.1
AND hAperMag3Err<=0.1
AND k\_1AperMag3Err<=0.1

\section{Membership probability calculation and distance estimation using PyUPMASK} \label{pyup}
Py-UPMASK begins by finding clusters in a multi-dimensional non-positional space, containing the proper-motions and the parallaxes. In this way, it also accounts (albeit indirectly) for clustering in the line-of-sight space. 
Once it has identified all the clusters in this space, it examines the positions of stars in those clusters, and rejects the ones that are consistent with a uniform distribution. This unique ability of the algorithm in simultaneously discerning clustering in the complete positional and proper motion space is the reason for opting it.

In order to examine potential sources of biases in the stellar data used for deriving distance to R1, we plot the stellar overdensity map of our Galaxy based on \emph{Gaia} DR3 \citep{poggio2021galactic} for a face-on orientation in the left panel of Fig~\ref{arms}. The red and blue areas correspond to over- and under-densities, respectively. The location of the Sun and the Galactic centre are marked as a yellow and black cross, respectively. The blue and red curves show the Saggitarius and Scutum arms, respectively -- as determined by \citet{reid2019trigonometric} using molecular masers around young, massive stars. The grey circles are the \emph{Gaia} stars located within R1.

This clearly shows that our stars are not localised in the line-of-sight space and rather span a vast range (near zero to about 12 kpc). As such, the sample is bound to contain stars that do not belong to the bubble under consideration. Note that the loci of spiral arms intersect with our stars’ trail, indicating that some of our selected stars belong to the two spiral arms. The stellar overdensity map also shows high overdensities in the area between these arms that is spanned by our stars. We find that the ensemble of ATLASGAL clumps in R1 have a minimum near-(kinematic) distance of 4.5 kpc. However, the spiral arms intercept our stars below 4 kpc (left panel), implying that stars at low distances in our sample are likely not associated with R1 and biasing our distance estimate to a lower value.

\begin{figure*}
\includegraphics[width=1\textwidth, trim= 0 0.1cm 0 0cm]{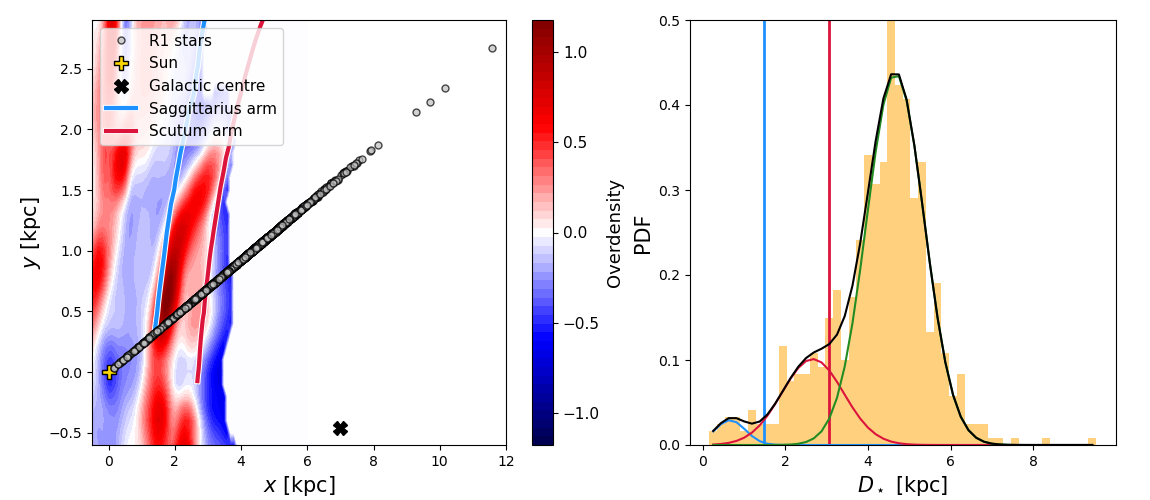} 
\caption{The left panel shows the face-on stellar overdensity map for the Milky Way based on the \emph{Gaia} DR3 \citep{poggio2021galactic}, where the overdensity decreases from red to blue. The Saggitarius and Scutum spiral arm models from \citet{reid2019trigonometric} are shown as blue and red curves, respectively. The Sun and the Galactic centre are shown as the yellow and black crosses, respectively. The \emph{Gaia} stars located within R1 in projection are shown as grey points. The right panel shows the probability density function of the trigonometric distances of the stars with clustering probability above 0.8. The blue and red vertical lines mark the distances where the Saggitarius and Scutum arm intersect our stellar trail in the left panel. The black curve is the best-fit multi-gaussian model for this distribution; the three Gaussians constituting the black model are shown in blue, red, and green.} \label{arms}
\end{figure*}

To check this, we plot the distance distribution of stars within R1 with clustering probabilities above 0.8 (right panel, Fig~\ref{arms}.). The rationale is that the stars associated with the spiral arms will exhibit a higher degree of clustering, and thus show up in the distribution. The vertical lines show the distances where the spiral arm models intercept the trail of stars within R1. The distribution clearly appears to be contained three peaks, with increasing levels from left to right. The best-fit multi-Gaussian model to this distribution is shown in black,
and is an excellent description of the data. The constituent Gaussians are shown in blue, red and green. Note that the positions of the peaks of blue and red distributions are close to those of the spiral arms. Since these stars possess a high degree of clustering, we infer that the distributions correspond to stars that are genuinely associated with the Saggittarius and Scutum arms, respectively. This leaves us with the third distribution in green, which we conclude to correspond to stars located in R1 within the bubble N59. We take its mean (4.66 kpc) and standard deviation (0.70 kpc) as R1’s distance and its uncertainty.

\bsp
\label{lastpage}
\end{document}